\documentclass{amsart}%
\usepackage{amsfonts}
\usepackage{amsmath}
\usepackage{amssymb}
\usepackage{graphicx}%
\usepackage{enumerate}
\newtheorem{theorem}{Theorem}[section]
\newtheorem{corollary}[theorem]{Corollary}
\newtheorem{lemma}[theorem]{Lemma}
\newtheorem{proposition}[theorem]{Proposition}
\theoremstyle{definition}
\newtheorem{definition}[theorem]{Definition}
\newtheorem{example}[theorem]{Example}

\newtheorem{remark}[theorem]{Remark}
\newtheorem{remarks}[theorem]{Remarks}
\theoremstyle{remark}
\newtheorem*{acknowledgements}{Acknowledgements}
\numberwithin{equation}{section}
\newlength{\equallabelwidth}
\begin{document}
\title{Harmonic analysis of iterated function systems with overlap}
\author{Palle E. T. Jorgensen}
\address[Palle E. T. Jorgensen]{Department of Mathematics, The University of Iowa, Iowa
City, IA 52242-1419, U.S.A.}
\email{jorgen@math.uiowa.edu}
\urladdr{http://www.math.uiowa.edu/\symbol{126}jorgen/}
\author{Keri A. Kornelson}
\author{Karen L. Shuman}
\address[Keri Kornelson, Karen Shuman]{Department of Mathematics and Statistics,
Grinnell College, Grinnell, IA 50112-1690, U.S.A.}
\urladdr{http://www.math.grinnell.edu/\symbol{126}kornelso/}
\urladdr{http://www.math.grinnell.edu/\symbol{126}shumank/}

\thanks{This material is based upon work supported by the U.S. National Science
Foundation under grants DMS-0457581 and DMS-0503990}
\dedicatory{Dedicated to the memory of Thomas P. Branson}\subjclass[2000]{33C50, 42C15, 46E22, 47B32}
\keywords{tight frames, Parseval frames, Bessel sequences, Hilbert space, polar
decomposition, attractor, IFS, equilibrium measure, composition operator}

\begin{abstract}
An iterated function system (IFS) is a system of contractive mappings
$\tau_{i} \colon Y \rightarrow Y $, $i = 1, \dots, N$ (finite) where $Y$ is a
complete metric space. Every such IFS has a unique (up to scale) equilibrium
measure (also called the Hutchinson measure $\mu$), and we study the Hilbert
space $L^{2}(\mu)$.

In this paper we extend previous work on IFSs without overlap. Our method
involves systems of operators generalizing the more familiar Cuntz relations
from operator algebra theory, and from subband filter operators in signal processing.

These Cuntz-like operator systems were used in recent papers on wavelets
analysis by Baggett, Jorgensen, Merrill and Packer, where they serve as a
first step to generating wavelet bases of Parseval type (alias normalized
tight frames), i.e., wavelet bases with redundancy.

Similarly, it was shown in work by Dutkay and Jorgensen that the iterative
operator approach works well for generating wavelets on fractals from IFSs
without overlap. But so far the more general and more difficult case of
essential overlap has resisted previous attempts at a harmonic analysis, and
explicit basis constructions in particular.

The operators generating the appropriate Cuntz relations are composition
operators, e.g., $F_{i}\colon f\rightarrow f\circ\tau_{i}$ where $(\tau_{i})$
is the given IFS. If the particular IFS is essentially non-overlapping, it is
relatively easy to compute the adjoint operators $S_{i}=F_{i}^{\ast}$, and the
$S_{i}$ operators will be isometries in $L^{2}(\mu)$ with orthogonal ranges.
For the case of essential overlap, we can use the extra terms entering in the
computation of the operators $F_{i}^{\ast}$ as a \textquotedblleft
measure\textquotedblright\ of the essential overlap for the particular IFS we
study. Here the adjoint operators $F_{i}^{\ast}$ refers to the Hilbert space
$L^{2}(\mu)$ where $\mu$ is the equilibrium measure $\mu$ for the given IFS
$(\tau_{i})$.

\end{abstract}
\maketitle

\tableofcontents

\section{\label{Int}Introduction: IFSs and operator theory}

In this paper we explore an operator theoretic approach
to self similarity and fractals which is common to problems involving
both iterated function systems (IFSs), and an aspect of quantum
communication.

Each of the two areas involves a finite set of
operations: in the case of IFSs they are geometric, and in the
quantum case, they involve channels of Hilbert spaces and associated
operator systems. The particular aspects of IFSs we have in mind are
studied in \cite{DuJo06d}; and the relevant results from quantum
communication in \cite{KLPL06} and \cite{Kri05}. We begin the Introduction with
some background on IFSs, and we motivate our present operator
theoretic approach. The operator theory and its applications are then
taken up more systematically in section 2 below.

It was proved recently that a class of non-linear fractals is amenable to a
computational harmonic analysis. While this work is motivated by applications,
it is significant in its own right; for example these fractals do not carry
the structure of groups, and so they do not come with a Haar measure.
Nonetheless, they have natural equilibrium measures $\mu$ which serve as
substitutes. Specifically, these fractals are known as iterated function
systems (IFS) indicating the recursive nature of their construction. But the
analysis so far has been limited to IFSs without essential overlap (see
Definition \ref{DefCon.EssOve}). The IFS fractals $X$ are built by passing to
a certain limit in a scaling iteration, and the \textquotedblleft
overlap\textquotedblright\ here refers to the parts of smaller scale-level. We
identify \textquotedblleft overlap\textquotedblright\ for a given $X$ relative
to the equilibrium measure $\mu$ for $X$. While there is prior work on some 
isolated cases of overlap, the cases that are understood are those of
negligible overlap. In this paper we identify and study the harmonic analysis
of IFSs with \textquotedblleft substantial overlap.\textquotedblright

An iterated function system (IFS) is a system of contractive mappings
$\tau_{i} \colon Y \rightarrow Y $, $i = 1, \dots, N$ (finite) where Y is a
complete metric space. Every such IFS has a unique (up to scale) equilibrium
measure (also called the Hutchinson measure $\mu$), and we study the Hilbert
space $L^{2}(\mu)$.

In this paper we extend previous work on IFSs without overlap. Our method
involves systems of operators generalizing the more familiar Cuntz relations
from operator algebra theory, and from subband filter operators in signal
processing. Before turning to the details, below we outline briefly the
operator-theoretic approach to IFSs.

For each $N$, there is a simple Cuntz $C^{\ast}$-algebra on generators and
relations, and its representations offer a useful harmonic analysis of general
IFSs, but there is a crucial difference between IFSs without overlap and those
with essential overlap (this will be made precise in Section \ref{Mul}). The
operators generating the appropriate Cuntz relations \cite{Cun77} are
composition operators, e.g., $F_{i}\colon f\rightarrow f\circ\tau_{i}$ where
$(\tau_{i})$ is the given IFS. If the particular IFS is essentially
non-overlapping, it is relatively easy to compute the adjoint operators
$S_{i}=F_{i}^{\ast}$, and the $S_{i}$ operators will be isometries in
$L^{2}(\mu)$ with orthogonal ranges. In a way, for the more difficult case of
essential overlap, we can use the extra terms entering into the computation of
the adjoint operators $F_{i}^{\ast}$ as a \textquotedblleft
measure\textquotedblright\ of the essential overlap for the particular IFS we
study. Here the adjoint operators $F_{i}^{\ast}$ refer to the Hilbert space
$L^{2}(\mu)$ where $\mu$ is carefully chosen. When the IFS is given, there are
special adapted measures $\mu$. We will be using the equilibrium measure $\mu$
for the given IFS $(\tau_{i})$, and by \cite{Hut81}, this $\mu$ contains much
essential information about the IFS, even in the classical cases of IFSs
coming from number theory.

So far, earlier work in the general area was restricted to IFSs without
overlap. For example, Cuntz-like operator systems were used in recent papers
on wavelets analysis by Baggett, Jorgensen, Merrill and Packer
\cite{BJMP04,BJMP05,BJMP06}, where they serve as a first step to generating
wavelet bases of Parseval type (alias normalized tight frames), i.e., wavelet
bases with redundancy.

Similarly, it was shown in work by Dutkay and Jorgensen \cite{DuJo06b} that
the iterative operator approach works well for generating wavelets on fractals
from IFSs without overlap. But so far the more general and more difficult case
of essential overlap has resisted previous attempts at a harmonic analysis,
and explicit basis constructions in particular.

In this paper (Section \ref{Ove}) we show that certain combinatorial cycles
from a symbolic encoding \cite{DuJo06e} of our IFSs yield an attractive
computational analysis of the ``IFS-overlap.'' Some results (e.g., Corollary
\ref{CorOveNov29.LowerBound}) are stated only for a model example, but as
indicated the idea with cycle-counting works in general.

While the modern study of iterated function systems (IFSs) has roots in
classical problems from number theory, infinite products \cite{Kol77}, and
more generally from probability and harmonic analysis (see, e.g.,
\cite{FLP94}), the subject took a more systematic turn in 1981 with the
influential paper \cite{Hut81} by Hutchinson. Since then, IFSs have found uses
in geometry (e.g., \cite{Bar06,BHS05}), in infinite network problems (e.g.,
\cite{GRS01}), in wavelets (e.g., \cite{BJMP05,BJMP06,Jor05a}), and in
dynamics and operator theory \cite{Kaw05,Jor06a,DuJo06a,DuJo06e,DuJo05b}.

But in operator algebras, the subject has its own independent start with the
paper by Cuntz \cite{Cun77}. In operator-algebra theory, a class of
representations of what have become known as the Cuntz, or the Cuntz--Krieger,
algebras is ideally suited for the study of iterative processes (including
IFSs) in mathematics and in mathematical physics. Perhaps the paper
\cite{CuKr80b} started this trend, but it was continued in various guises in
the following papers: see \cite{Pop89} and later related papers by Popescu,
\cite{BaVi05b}, \cite{JoPe96}, \cite{JoKr03}, and \cite{Jor04a}, among others.
The Cuntz and the related $C^{*}$-algebras have become ubiquitous tools in the
study of iterative systems, or in the analysis of fractals or of graphs. One
reason for this is that there is an unexpected connection to signal and image
processing; see \cite{Jor06a}. They are algebras on generators and relations;
and often, as is the case in the present paper, the generators may be
identified as substitution operators \cite{Jor04a}. However, these
substitution operators and their adjoints (often called transfer operators)
are of independent interest; see, e.g., \cite{JoPe96,Sin93,KuSh93,Kwa04}.

Since the Cuntz and related $C^{*}$-algebras typically are naturally generated
by a finite set of concrete operators, it has become customary to study such
finite sets as matrices with operator entries; for example, a class of vectors
with operator entries are studied in \cite{Arv04} as row contractions. In
fact, it is convenient to distinguish between row vectors with operator
entries, and the corresponding columns of operators: one is the adjoint of the
other. In this paper, our study of IFSs with overlap leads naturally to a
family of column isometries, where the operator entries in our column vectors
are substitution operators built from the maps that define the IFS under consideration.

While there is already a rich literature on IFSs without overlap, the more
difficult case of overlap has received relatively less attention; see,
however, \cite{Bar06,FLP94,Sol95,Sol98}. The point of view of the latter three
papers is generalized number expansions in the form of random variables: Real
numbers are expanded in a basis which is a fraction, although the ``digits''
are bits; with infinite strings of bits identified in a Bernoulli probability
space. It turns out the distribution of the resulting random variables is
governed by the measures which arise as a special cases of Hutchinson's
analysis \cite{Hut81} of IFSs with overlap. See also \cite{HuRu00}. However,
concrete results about these measures have been elusive. For example, it is
proved in \cite{Sol95} that the measures for expansions in a basis
corresponding to IFSs with overlap and given by a scaling parameter are known
to be absolutely continuous for a.e.\ value of the parameter.

In this paper we consider a rather general class of IFSs with overlap. We show
that they can be understood in terms of the spectral theory of Cuntz-like
column isometries. Moreover we show that our column isometries yield exact
representations of the Cuntz relations precisely when the IFS has overlap of
measure zero, where the measure is an equilibrium measure $\mu$ of the
Hutchinson type.

There is a separate development related to the study of Fourier bases in the
Hilbert spaces $L^{2}(\mu)$ for $\mu$ an IFS equilibrium measure; see
\cite{JoPe98b,Str00,StUs00,LaWa00,Jor06a,DuJo06e}. We address some of these
issues below, but it turns out that exact Fourier bases for the case of IFS
with overlap are harder to come by. Further, in \cite{DuJo06e}, the coauthors
noted that Fourier bases in $L^{2}(\mu)$ can only exist if the equilibrium
measure $\mu$ has equal weights, and so in the present paper, we make this assumption.

Recent references \cite{DuJo05b,DuJo06a,DuJo06b,DuJo06e,Jor06a} deal with
wavelet constructions on non-linear attractors $X$, such as arise from systems
of branched mappings, e.g., finite affine systems (affine IFSs), or Julia sets
\cite{Bea91} generated by branches of the inverses of given rational mappings
of one complex variable. These $X$ come with associated equilibrium measures
$\mu$. However, due to geometric obstructions, it is typically not possible
for the corresponding $L^{2}(X, \mu)$ to carry orthonormal bases (ONBs) of
complex exponentials, i.e., Fourier bases. We showed instead in \cite{DuJo06b}
that wavelet bases may be constructed in ambient $L^{2}(\mu)$ Hilbert spaces
via multiresolutions and representations of Cuntz-like operator relations in
$L^{2}(X, \mu)$ as in Definitions \ref{DefMul.ColIso} and \ref{DefMul.RepCun}
in the next section.

\section{\label{Mul}Multivariable operator theory}

        In quantum communication (the study of (quantum)
error-correction codes), certain algebras of operators and completely
positive mappings form the starting point; see especially the papers
\cite{KLPL06} and \cite{Kri05}. They take the form of a finite number of
channels of Hilbert space operators $F_i$ which are assumed to satisfy
certain compatibility conditions. The essential one is that the
operators from a partition of unity, or rather a partition of the
identity operator $I$ in the chosen Hilbert space. Here (Definition \ref{DefMul.ColIso}) we call such a system $(F_i)$ a column isometry. An extreme case
of this is when a certain Cuntz relation (Definition \ref{DefMul.RepCun}) is
satisfied by $(F_i)$. Referring back to our IFS application, the
extreme case of the operator relations turn out to correspond to the
limiting case of non-overlap, i.e., to the case when our IFSs have no
essential overlap (Definition \ref{DefCon.EssOve}).

In this section we outline some uses of ideas from multivariable operator
theory (see, e.g., \cite{Arv04}) in iterated function systems (IFS) with
emphasis on \emph{IFSs with overlap}. The tools we use are \emph{column
isometries} and systems of \emph{composition operators}.

\begin{definition}
\label{DefMul.ColIso}Let $\mathcal{H}$ be a complex Hilbert space, and let
$N\in\mathbb{N}$, $N\geq2$. A system $\left(  F_{1},\dots,F_{N}\right)  $ of
bounded operators in $\mathcal{H}$ is said to be a \emph{column isometry} if
the mapping%
\begin{equation}
\mathbb{F}\colon\mathcal{H}\longrightarrow%
\begin{pmatrix}
\mathcal{H}\\
\oplus\\
\vdots\\
\oplus\\
\mathcal{H}%
\end{pmatrix}
:\xi\longmapsto%
\begin{pmatrix}
F_{1}\xi\\
\vdots\\
F_{N}\xi
\end{pmatrix}
\label{eqMul.1}%
\end{equation}
is isometric. Here we write the $N$-fold orthogonal sum of $\mathcal{H}$ in
column form, but we will also use the shorter notation $\mathcal{H}_{N}$. As a
Hilbert space, it is the same as $\mathcal{H}\oplus\dots\oplus\mathcal{H}$,
but for clarity it is convenient to identify the \emph{adjoint operator}
$\mathbb{F}^{\ast}$ as a row%
\begin{equation}
\mathbb{F}^{\ast}\colon\underbrace{\mathcal{H}\oplus\dots\oplus\mathcal{H}%
}_{N\text{ times}}\longrightarrow\mathcal{H}:\left(  \xi_{1},\dots,\xi
_{N}\right)  \longrightarrow\sum_{i=1}^{N}F_{i}^{\ast}\xi_{i}. \label{eqMul.2}%
\end{equation}
The inner product in $\mathcal{H}_{N}$ is $\sum_{i=1}^{N}\left\langle
\,\xi_{1}\mid\eta_{i}\,\right\rangle $, and relative to the respective inner
products on $\mathcal{H}_{N}$ and on $\mathcal{H}$, we have%
\begin{equation}
\left\langle
\,\smash{\underset{\overbrace{\begin{smallmatrix}\text{column} \\ \text{operator}\end{smallmatrix}}}{\mathbb{F}}}\xi
\Biggm|\smash{\begin{pmatrix}\eta _{1} \\ \vdots \\ \eta _{N}\end{pmatrix}}\,\right\rangle
_{\mathcal{H}_{N}}=\left\langle \,\xi
\Biggm|\smash{\underset{\overbrace{\begin{smallmatrix}\text{row}\vphantom{\text{column}} \\ \text{operator}\end{smallmatrix}}}{\phantom{{}^{\ast }}\mathbb{F}^{\ast }}\begin{pmatrix}\eta _{1} \\ \vdots \\ \eta _{N}\end{pmatrix}}\,\right\rangle
_{\mathcal{H}}%
\vphantom{\underset{\overbrace{\begin{smallmatrix}\text{column} \\ \text{operator}\end{smallmatrix}}}{\mathbb{F}}\begin{pmatrix}\eta _{1} \\ \vdots \\ \eta _{N}\end{pmatrix}}.
\label{eqMul.3}%
\end{equation}

\end{definition}

\begin{remark}
\label{RemMul.PropProj}In the definition of a column isometry, it is stated
that $\mathbb{F}$ in \textup{(\ref{eqMul.1})} is \emph{isometric}
$\mathcal{H}\rightarrow\mathcal{H}_{N}$, but it is not necessarily onto
$\mathcal{H}_{N}$. This means that in general the matrix operator
$\mathbb{FF}^{\ast}\colon\mathcal{H}_{N}\rightarrow\mathcal{H}_{N}$ is a
proper \emph{projection}. By this we mean that the block matrix%
\begin{equation}
\mathbb{FF}^{\ast}=\left(  F_{i}F_{j}^{\ast}\right)  _{i,j=1}^{N}
\label{eqMul.3bis}%
\end{equation}
satisfies the following system of identities:%
\begin{equation}
\sum_{k=1}^{N}\left(  F_{i}F_{k}^{\ast}\right)  \left(  F_{k}F_{j}^{\ast
}\right)  =F_{i}F_{j}^{\ast},\qquad1\leq i,j\leq N. \label{eqMul.4}%
\end{equation}
Note that $\mathbb{FF}^{\ast}=I_{\mathcal{H}_{N}}$ if and only if%
\begin{equation}
F_{i}F_{j}^{\ast}=\delta_{i,j}I,\qquad1\leq i,j\leq N. \label{eqMul.5}%
\end{equation}

\end{remark}

\begin{definition}
\label{DefMul.RepCun}A column isometry $\mathbb{F}$ satisfies $\mathbb{FF}%
^{\ast}=I_{\mathcal{H}_{N}}$ if and only if it defines a representation of the
Cuntz algebra $\mathcal{O}_{N}$. In that case, the operators $S_{i}%
:=F_{i}^{\ast}$ are isometries in $\mathcal{H}$ with orthogonal ranges, and%
\begin{equation}
\sum_{i=1}^{N}S_{i}S_{i}^{\ast}=I_{\mathcal{H}}. \label{eqMul.6}%
\end{equation}

\end{definition}

\begin{remark}
\label{RemMul.DimFin}The distinction between the operator relations in
Definitions \textup{\ref{DefMul.ColIso}} and \textup{\ref{DefMul.RepCun}} is
much more than a technicality: Definition \textup{\ref{DefMul.RepCun}} is the
more restrictive. Because of the orthogonality axiom in Definition
\textup{\ref{DefMul.RepCun}}, it is easy to see that if a Hilbert space
$\mathcal{H}$ carries a nonzero representation of the Cuntz relations
\textup{(}Definition \textup{\ref{DefMul.RepCun}) (}see \cite{Cun77}%
\textup{)}, then $\mathcal{H}$ must be infinite-dimensional, reflecting the
infinitely iterated and orthogonal subdivision of projections, a hallmark of fractals.

In contrast, the condition of Definition \textup{\ref{DefMul.ColIso}}, or
equivalently $\sum_{i=1}^{N}F_{i}^{\ast}F_{i}=I_{\mathcal{H}}$, may easily be
realized when the dimension of the Hilbert space $\mathcal{H}$ is finite. In
fact such representations are used in quantum computation; see, e.g.,
\cite[Theorem 2]{LiSe05} and \cite{Kri05}.
\end{remark}

\begin{definition}
\label{DefMul.AbsCon}Let $\left(  X,\mathcal{B},\mu\right)  $ be a finite
measure space, i.e., $X$ is a set, $\mathcal{B}$ is a sigma-algebra of subsets
in $X$, and $\mu$ is a finite positive measure defined on $\mathcal{B}$. We
will assume that $\mathcal{B}$ is complete, and the Hilbert space
$L^{2}\left(  X,\mathcal{B},\mu\right)  $ will be denoted $L^{2}\left(
\mu\right)  $ for short. If $\nu$ is a second measure, also defined on
$\mathcal{B}$, we say that $\nu\ll\mu$ \textup{(}relative absolute
continuity\/\textup{) if}%
\begin{equation}
S\in\mathcal{B},\qquad\mu\left(  S\right)  =0\implies\nu\left(  S\right)  =0.
\label{eqMul.7}%
\end{equation}
In that case, the corresponding Radon--Nikodym derivative will be denoted
$\varphi:=\frac{d\nu}{d\mu}$; i.e., we have $\varphi\in L^{1}\left(
\mu\right)  $, and
\begin{equation}
\nu\left(  S\right)  =\int_{S}\varphi\left(  x\right)  \,d\mu\left(  x\right)
\text{\qquad for all }S\in\mathcal{B}. \label{eqMul.8}%
\end{equation}

An endomorphism $\tau\colon X\rightarrow X$ is said to be \emph{measurable} if%
\begin{equation}
S\in\mathcal{B}\implies\tau^{-1}\left(  S\right)  \in\mathcal{B},
\label{eqMul.9}%
\end{equation}
where $\tau^{-1}\left(  S\right)  =\left\{  \,x\in X\mid\tau\left(  x\right)
\in S\,\right\}  $. In that case, a measure $\mu\circ\tau^{-1}$ is
\emph{induced} on $\mathcal{B}$, and given by
\begin{equation}
\left(  \mu\circ\tau^{-1}\right)  \left(  S\right)  :=\mu\left(  \tau
^{-1}\left(  S\right)  \right)  ,\qquad S\in\mathcal{B}. \label{eqMul.10}%
\end{equation}

\end{definition}

\begin{remark}
\label{RemMul.ComOp}Every measurable endomorphism $\tau\colon X\rightarrow X
$, referring to $\left(  X,\mathcal{B},\mu\right)  $, induces a
\emph{composition operator}%
\begin{equation}
C_{\tau}\colon L^{2}\left(  \mu\right)  \ni f\longmapsto f\circ\tau\in
L^{2}\left(  \mu\right)  , \label{eqMul.11}%
\end{equation}
and it can be shown that $C_{\tau}$ is a \emph{bounded} operator in the
Hilbert space $L^{2}\left(  \mu\right)  $ if and only if $\mu\circ\tau^{-1}%
\ll\mu$ with Radon--Nikodym derivative in $L^{\infty}$. Moreover, if $\mu
\circ\tau^{-1}\ll\mu$, set $\varphi:=\frac{d\mu\circ\tau^{-1}}{d\mu}$.
\end{remark}

\begin{definition}
\label{DefMul.FinTyp}A measurable endomorphism $\tau\colon X\rightarrow X$ is
said to be of \emph{finite type} if there are a \emph{finite partition}
$E_{1},\dots,E_{k}$ of $\tau\left(  X\right)  $ and measurable mappings
$\sigma_{i}\colon E_{i}\rightarrow X$, $i=1,\dots,k$ such that
\begin{equation}
\sigma_{i}\circ\tau|_{E_{i}}=\operatorname{id}_{E_{i}},\qquad1\leq i\leq k.
\label{eqMul.12}%
\end{equation}

\end{definition}

\begin{lemma}
\label{LemMul.AdjForm}Let $\left(  X,\mathcal{B},\mu\right)  $ be as above,
and let $\tau\colon X\rightarrow X$ be a measurable endomorphism of finite
type. Suppose $\mu\circ\tau^{-1}\ll\mu$, and set $\varphi:=\frac{d\mu\circ
\tau^{-1}}{d\mu}$.\renewcommand{\theenumi}{\alph{enumi}}

\begin{enumerate}
\item \label{LemMul.AdjForm(1)}Then $\varphi$ is supported \textup{(}$\mu
$-$\mathrm{a.e.}$\textup{)} on $\tau\left(  X\right)  .$

\item \label{LemMul.AdjForm(2)}Moreover, if $E_{1},\dots,E_{k}$ is a
\textup{(}non-overlapping\/\textup{)} partition as in \textup{(\ref{eqMul.12}%
)}, then the adjoint operator $C_{\tau}^{\ast}$ of the composition operator
\textup{(\ref{eqMul.12})} is given by the formula%
\begin{equation}
C_{\tau}^{\ast}f|_{E_{i}}=\varphi\left(  f\circ\sigma_{i}\right)  ,\qquad
i=1,\dots,k,\;f\in L^{2}\left(  \mu\right)  ; \label{eqMul.13}%
\end{equation}
i.e., for $x\in E_{i}$, $\left(  C_{\tau}^{\ast}f\right)  \left(  x\right)
=\varphi\left(  x\right)  f\left(  \sigma_{i}\left(  x\right)  \right)  $.
\end{enumerate}
\end{lemma}

\begin{proof}
We will begin by assuming $\mu\circ\tau^{-1}\ll\mu$. At the end of the
resulting computation, it will then be clear that the reasoning is in fact
reversible. The composition operator $C_{\tau}$ is well defined as in
(\ref{eqMul.12}), referring to the Hilbert space $L^{2}\left(  \mu\right)  $
with its usual inner product%
\begin{equation}
\left\langle \,f_{1}\mid f_{2}\,\right\rangle _{\mu}:=\int_{X}\overline{f_{1}%
}\,f_{2}\,d\mu. \label{eqMul.14}%
\end{equation}
For the adjoint operator $C_{\tau}^{\ast}$, we have%
\begin{align*}
\left\langle \,C_{\tau}^{\ast}f_{1}\mid f_{2}\,\right\rangle _{\mu}  &
=\left\langle \,f_{1}\mid C_{\tau}f_{2}\,\right\rangle _{\mu}=\left\langle
\,f_{1}\mid f_{2}\circ\tau\,\right\rangle _{\mu}\\
&  =\int_{X}\overline{f_{1}}\,\left(  f_{2}\circ\tau\right)  \,d\mu=\sum
_{i=1}^{k}\int_{\tau^{-1}\left(  E_{i}\right)  }\overline{f_{1}}\,\left(
f_{2}\circ\tau\right)  \,d\mu\\
&  =\sum_{i=1}^{k}\int_{\tau^{-1}\left(  E_{i}\right)  }\left(  \overline
{f_{1}}\circ\sigma_{i}\circ\tau\right)  \left(  f_{2}\circ\tau\right)
\,d\mu\\
&  =\sum_{i=1}^{k}\int_{E_{i}}\left(  \overline{f_{1}}\circ\sigma_{i}\right)
\,f_{2}\,d\mu\circ\tau^{-1}\\
&  =\sum_{i=1}^{k}\int_{E_{i}}\left(  \overline{f_{1}}\circ\sigma_{i}\right)
\,f_{2}\,\varphi\,d\mu
\end{align*}
for all $f_{1},f_{2}\in L^{2}\left(  \mu\right)  $. The desired formula
(\ref{eqMul.13}) for the adjoint operator $C_{\tau}^{\ast}$ follows.
\end{proof}

\begin{definition}
\label{DefMul.IFS}Let $\left(  X,\mathcal{B},\mu\right)  $ be a finite measure
space. A system $\tau_{1},\dots,\tau_{N}$ of measurable endomorphisms is said
to be an \emph{iterated function system} \textup{(}\emph{IFS\/}\textup{)}. Let
$p_{i}\in\left[  \,0,1\,\right]  $ be given such that $\sum_{i=1}^{N}p_{i}=1$.
If%
\begin{equation}
\sum_{i=1}^{N}p_{i}\mu\circ\tau_{i}^{-1}=\mu\label{eqMul.15}%
\end{equation}
we say that the measure $\mu$ is a $p$\emph{-equilibrium} measure. Motivated
by earlier work on the harmonic analysis of equilibrium measures
\textup{(}see, e.g., \cite{JoPe98b} and \cite{DuJo06e}\textup{)}, we will here
restrict attention to the case of equal weights, i.e., assume that
$p_{1}=\dots=p_{N}=1/N$; and we shall then refer to $\mu$ as an
\emph{equilibrium measure}, with the understanding that $p_{i}=1/N$. In
general, equilibrium measures might not exist; and even if they do, they may
not be unique.
\end{definition}

\begin{proposition}
\label{ProMul.EquMea}Let $\left(  X,\mathcal{B},\mu\right)  $ be a finite
measure space, and let $\tau_{1},\dots,\tau_{N}$ be measurable endomorphisms.
Then some $\mu$ is an equilibrium measure if and only if the associated linear
operator%
\begin{equation}
\mathbb{F}_{\tau}\colon L^{2}\left(  \mu\right)  \longrightarrow%
\begin{pmatrix}
L^{2}\left(  \mu\right) \\
\oplus\\
\vdots\\
\oplus\\
L^{2}\left(  \mu\right)
\end{pmatrix}
:f\longmapsto\frac{1}{\sqrt{N}}%
\begin{pmatrix}
f\circ\tau_{1}\\
\vdots\\
f\circ\tau_{N}%
\end{pmatrix}
\label{eqMul.16}%
\end{equation}
is isometric, i.e., if and only if the individual operators
\begin{equation}
F_{i}\colon f\longmapsto\frac{1}{\sqrt{N}}f\circ\tau_{i}\text{\qquad in }%
L^{2}\left(  \mu\right)  \label{eqMul.17}%
\end{equation}
define a column isometry.
\end{proposition}

\begin{proof}
Using polarization for the inner product $\left\langle \,\,\cdot\mid
\cdot\,\right\rangle _{\mu}$ in (\ref{eqMul.14}), we first note that $\mu$ is
an equilibrium measure if and only if%
\begin{equation}
\frac{1}{N}\sum_{i=1}^{N}\int_{X}\left\vert f\right\vert ^{2}\circ\tau
_{i}\,d\mu=\int_{X}\left\vert f\right\vert ^{2}\,d\mu\qquad(=:\left\Vert
f\right\Vert _{L^{2}\left(  \mu\right)  }^{2}) \label{eqMul.18}%
\end{equation}
holds for all $f\in L^{2}\left(  \mu\right)  $.

The terms on the left-hand side in (\ref{eqMul.18}) are $\frac{1}{N}\int
_{X}\left\vert f\right\vert ^{2}\circ\tau_{i}\,d\mu=\left\Vert F_{i}%
f\right\Vert _{L^{2}\left(  \mu\right)  }^{2}$, so (\ref{eqMul.18}) is
equivalent to%
\[
\left\langle \,f\Bigm|\smash{\sum_{i=1}^{N}F_{i}^{\ast }F_{i}f}\,\right\rangle
_{\mu}\vphantom{\sum_{i=1}^{N}F_{i}^{\ast }F_{i}f}=\sum_{i=1}^{N}\left\langle
\,F_{i}f\mid F_{i}f\,\right\rangle =\sum_{i=1}^{N}\left\Vert F_{i}f\right\Vert
_{L^{2}\left(  \mu\right)  }^{2}=\left\Vert f\right\Vert _{L^{2}\left(
\mu\right)  }^{2},
\]
which in turn is the desired operator identity%
\begin{equation}
\sum_{i=1}^{N}F_{i}^{\ast}F_{i}=I_{L^{2}\left(  \mu\right)  } \label{eqMul.19}%
\end{equation}
that defines $\mathbb{F}$ as a column isometry.
\end{proof}

Our main result in Sections \ref{Con} and \ref{Ove} will yield a formula for
the Radon--Nikodym derivatives $\varphi_{i}:=\frac{d\mu\circ\tau_{i}^{-1}%
}{d\mu}$; see Remark \ref{RemCon.EquWei} for details.

\section{\label{Con}Contractive iterated function systems}

The study of contractive iterated function systems (contractive IFSs) was
initiated in a systematic form by Hutchinson in \cite{Hut81}, but was also
used in harmonic analysis before 1981.

Suppose $N\in\mathbb{N}$, $N\geq2$, is given, and suppose some mappings
$\tau_{i}\colon Y\rightarrow Y$, $i=1,\dots,N$ are contractive in some
complete metric space $\left(  Y,d\right)  $: then we say that $\left(
\tau_{i}\right)  _{i=1}^{N}$ is a \emph{contractive IFS}. The contractivity
entails constants $c_{1},\dots,c_{N}$, $c_{i}\in\left(  0,1\right)  $, such
that%
\begin{equation}
d\left(  \tau_{i}\left(  x\right)  ,\tau_{i}\left(  y\right)  \right)  \leq
c_{i}d\left(  x,y\right)  ,\qquad x,y\in Y. \label{eqCon.1}%
\end{equation}

Hutchinson's first theorem \cite{Hut81} states that there is then a unique
compact subset $X\subset Y$ such that%
\begin{equation}
X=\bigcup_{i=1}^{N}\tau_{i}\left(  X\right)  . \label{eqCon.2}%
\end{equation}
This set $X$ is called the \emph{attractor} for the system.

If weights $\left(  p_{i}\right)  _{i=1}^{N}$ are given as in Definition
\ref{DefMul.IFS}, Hutchinson's second theorem states that there is a unique
(up to scale) positive finite (nonzero) Borel measure $\mu$ on $Y$ such that%
\begin{equation}
\mu=\sum_{i=1}^{N}p_{i}\mu\circ\tau_{i}^{-1}. \label{eqCon.3}%
\end{equation}
Moreover, if $p_{i}>0$ for all $i$, then the support of $\mu$ is the compact
set $X$ (fractal) in (\ref{eqCon.2}), the attractor.

For many purposes, it is convenient to normalize the measure $\mu$ in
(\ref{eqCon.3}) such that $\mu\left(  X\right)  =\mu\left(  Y\right)  =1$.
Because of applications to harmonic analysis \cite{DuJo06e}, we will also
restrict the weights $\left(  p_{i}\right)  $ in (\ref{eqCon.3}) such that
$p_{i}=1/N$.

\begin{definition}
\label{DefCon.LimMea}For Borel measures $\nu$ on $Y$ \textup{(}some given
complete metric space\/\textup{)}, set
\begin{equation}
T\nu:=\frac{1}{N}\sum_{i=1}^{N}\nu\circ\tau_{i}^{-1}. \label{eqCon.4}%
\end{equation}
Set%
\[
\operatorname{Lip}_{1}\left(  Y\right)  =\left\{  \,f:Y\rightarrow
\mathbb{R},\;\left\vert f\left(  x\right)  -f\left(  y\right)  \right\vert
\leq d\left(  x,y\right)  \,\right\}  ,
\]
and%
\begin{equation}
d_{1}\left(  \nu_{1},\nu_{2}\right)  :=\sup\left\{  \,\int f\,d\nu_{1}-\int
f\,d\nu_{2}\biggm|f\in\operatorname{Lip}_{1}\left(  Y\right)  \,\right\}  .
\label{eqCon.5}%
\end{equation}
Then it is easy to see \cite{Hut81} that $d_{1}$ is a metric and that the
probability measures form a complete metric space with respect to $d_{1}$.

Moreover, regardless of the choice of the initial measure $\nu$ \textup{(}some
probability measure on $Y$\textup{)}, the limit%
\begin{equation}
\lim_{n\rightarrow\infty}T^{n}\nu=\mu\text{\qquad\textup{(}in the metric
}d_{1}\text{\textup{)}} \label{eqCon.6}%
\end{equation}
exists, where $\mu$ is the equilibrium measure from \textup{(\ref{eqCon.3})},
$p_{i}=1/N$, and where the convergence in \textup{(\ref{eqCon.6})} is relative
to the metric $d_{1}$ from \textup{(\ref{eqCon.5})}.
\end{definition}

\begin{example}
\label{ExaCon.IntSin}Let $\mathbb{Z}_{N}:=\left\{  0,1,\dots,N-1\right\}  $,
and set $\Omega=\mathbb{Z}_{N}\times\mathbb{Z}_{N}\times\cdots=\left(
\mathbb{Z}_{N}\right)  ^{\mathbb{N}}$, i.e., the infinite Cartesian product.
We equip $\Omega$ with its usual Tychonoff topology, and its usual metric.
Points in $\Omega$ are denoted $\omega=\left(  \omega_{1}\omega_{2}%
\dots\right)  $, $\omega_{i}\in\mathbb{Z}_{N}$, $i=1,2,\dots$. For
$n\in\mathbb{N}$, set $\omega|n=\left(  \omega_{1}\omega_{2}\dots\omega
_{n}\right)  $. Further, we shall need the one-sided shifts%
\begin{align}
\sigma_{j}\left(  \omega_{1}\omega_{2}\dots\right)   &  =\left(  j\omega
_{1}\omega_{2}\dots\right)  ,\qquad j\in\mathbb{Z}_{N},\;\omega\in
\Omega,\label{eqCon.7}\\%
\intertext{and}%
\sigma\left(  \omega_{1}\omega_{2}\omega_{3}\dots\right)   &  =\left(
\omega_{2}\omega_{3}\dots\right)  ,\qquad\omega\in\Omega. \label{eqCon.8}%
\end{align}

If $\left(  \tau_{i}\right)  _{i=1}^{N}$ is a contractive IFS with attractor
$X$, it is easy to see that for each $\omega\in\Omega$, the intersection%
\begin{equation}
\bigcap_{n=1}^{\infty}\tau_{\omega|n}\left(  X\right)  \label{eqCon.9}%
\end{equation}
is a singleton. Here we use the notation%
\begin{equation}
\tau_{\omega|n}:=\tau_{\omega_{1}}\tau_{\omega_{2}}\cdots\tau_{\omega_{n}}.
\label{eqCon.10}%
\end{equation}

If $\omega\in\Omega$ is given, let $\pi\left(  \omega\right)  $ be the
\textup{(}unique\/\textup{)} point in the intersection \textup{(\ref{eqCon.9}%
)}. The mapping $\pi\colon\Omega\rightarrow X$ is called the \emph{encoding}.
\end{example}

\begin{lemma}
\label{LemCon.CodCon}Let $\left(  \tau_{i}\right)  _{i=1}^{N}$ and $X$ be as
above, i.e., assumed contractive. Then the coding mapping $\pi\colon
\Omega\rightarrow X$ from \textup{(\ref{eqCon.9})} is continuous, and we have%
\begin{equation}
\pi\circ\sigma_{j}=\tau_{j}\circ\pi,\qquad j=1,\dots,N\text{\quad or\quad}%
j\in\mathbb{Z}_{N}. \label{eqCon.11}%
\end{equation}

\end{lemma}

\begin{proof}
The continuity is clear from the definitions. We verify (\ref{eqCon.11}): Let
$\omega=\left(  \omega_{1}\omega_{2}\dots\right)  \in\Omega$. Then%
\begin{align*}
\left(  \pi\circ\sigma_{j}\right)  \left(  \omega\right)   &  =\pi\left(
j\omega_{1}\omega_{2}\dots\right)
=\smash{\bigcap_{n}\tau _{j}\tau _{\omega _{1}}\cdots \tau _{\omega _{n}}\left( X\right) }\\
&  =\tau_{j}\left(  \,\bigcap_{n}\tau_{\omega_{1}}\tau_{\omega_{2}}\cdots
\tau_{\omega_{n}}\left(  X\right)  \right) \\
&  =\tau_{j}\left(  \pi\left(  \omega\right)  \right)  =\left(  \tau_{j}%
\circ\pi\right)  \left(  \omega\right)  ,
\end{align*}
where we used contractivity of the mappings $\tau_{j}$.
\end{proof}

\begin{corollary}
\label{CorCon.BernMe}It follows from Hutchinson's theorem \cite{Hut81} that
for each $\left(  p_{1},\dots,p_{N}\right)  $ with $\sum_{1}^{N}p_{i}=1$,
$p_{i}\geq0$, there is a unique probability measure $P_{\left(  p\right)  }$
on $\Omega$ such that%
\begin{equation}
P_{\left(  p\right)  }=\sum_{i=1}^{N}p_{i}P_{\left(  p\right)  }\circ
\sigma_{i}^{-1}. \label{eqCon.12}%
\end{equation}
When $p_{i}=1/N$, this measure is called \emph{the Bernoulli measure} on
$\Omega$. The measures $P_{\left(  p\right)  }$ are also familiar
infinite-product measures, considered first by Kolmogorov \cite{Kol77}.
\end{corollary}

\begin{corollary}
\label{CorCon.EquiMe}Let $\left(  \tau_{i}\right)  _{i=1}^{N}$ be a
contractive IFS, and let $\left(  p_{i}\right)  $ be given such that
$\sum_{i=1}^{N}p_{i}=1$, $p_{i}\geq0$. Let $\pi\colon\Omega\rightarrow X$ be
the corresponding endoding mapping. Then the equilibrium measure $\mu_{\left(
p\right)  }$ for $\left(  \tau_{i}\right)  $ is%
\begin{equation}
\mu_{\left(  p\right)  }=P_{\left(  p\right)  }\circ\pi^{-1}, \label{eqCon.13}%
\end{equation}
i.e., for Borel subsets $E\subset X$ we have%
\begin{equation}
\mu_{\left(  p\right)  }\left(  E\right)  =P_{\left(  p\right)  }\left(
\pi^{-1}\left(  E\right)  \right)  , \label{eqCon.14}%
\end{equation}
where
\begin{equation}
\pi^{-1}\left(  E\right)  =\left\{  \,\omega\in\Omega\mid\pi\left(
\omega\right)  \in E\,\right\}  . \label{eqCon.15}%
\end{equation}

\end{corollary}

\begin{proof}
Let $\left(  \tau_{i}\right)  $, $\left(  p\right)  $, $\Omega$, and
$P_{\left(  p\right)  }$ be as described. The following computation uses
(\ref{eqCon.11}) in the form
\[
\sigma_{j}^{-1}\pi^{-1}=\pi^{-1}\tau_{j}^{-1},\qquad j\in\mathbb{Z}_{N}.
\]
The conclusion will follow from the uniqueness part of Hutchinson's theorem if
we check that $P_{\left(  p\right)  }\circ\pi^{-1}$ satisfies the identity
(\ref{eqMul.15}).

Let $E$ be a Borel set $\subset X$. Then%
\begin{align*}
\left(  P_{\left(  p\right)  }\circ\pi^{-1}\right)  \left(  E\right)   &
=P_{\left(  p\right)  }\left(  \pi^{-1}\left(  E\right)  \right) \\
&  =\sum_{i=1}^{N}p_{i}P_{\left(  p\right)  }\left(  \sigma_{i}^{-1}\left(
\pi^{-1}\left(  E\right)  \right)  \right) \\
&  =\sum_{i=1}^{N}p_{i}P_{\left(  p\right)  }\left(  \pi^{-1}\left(  \tau
_{i}^{-1}\left(  E\right)  \right)  \right) \\
&  =\sum_{i=1}^{N}p_{i}\left(  P_{\left(  p\right)  }\circ\pi^{-1}\right)
\circ\tau_{i}^{-1}\left(  E\right)  .
\end{align*}
Since $P_{\left(  p\right)  }\circ\pi^{-1}$ is a probability measure, it
follows that $\mu_{\left(  p\right)  }:=P_{\left(  p\right)  }\circ\pi^{-1}$
is the unique solution to (\ref{eqMul.15}) given by the normalization
$\mu_{\left(  p\right)  }\left(  X\right)  =1$.
\end{proof}

\begin{remark}
\label{RemCon.EquWei}It is immediate from the definitions that all the
measures $\mu$ solving \textup{(\ref{eqCon.3})} for the case of positive
weights $p_{i}>0$ will satisfy the relative absolute continuity condition%
\begin{equation}
\mu\circ\tau_{i}^{-1}\ll\mu, \label{eqRemCon.EquWeipound}%
\end{equation}
but we shall be especially interested in the case of equal weights $p_{i}%
=1/N$, and in this case we set
\begin{equation}
\varphi_{i}:=\frac{d\mu\circ\tau_{i}^{-1}}{d\mu}.
\label{eqRemCon.EquWeipoundpound}%
\end{equation}

\end{remark}

\begin{example}
\label{ExaCon.xplus1}Let $N=2$, $\mathbb{Z}_{2}=\left\{  0,1\right\}  $,
$\Omega=\left(  \mathbb{Z}_{2}\right)  ^{\mathbb{N}}$, $p_{0}=p_{1}=1/2$, and
let $\lambda\in\left(  0,1\right)  $ be given. Consider the two contractive
maps
\[
\tau_{0}\left(  x\right)  =\lambda x\text{\quad and\quad}\tau_{1}%
x=\lambda\left(  x+1\right)  \text{\qquad in }\mathbb{R}.
\]
For the attractor $X_{\lambda}$, we have the following three cases:

\textsc{Case 1:} $\lambda\in\left(  0,1/2\right)  $:
\begin{equation}
X_{\lambda}\text{ is a fractal of fractal dimension }D_{\lambda}=-\log
2/\log\lambda. \label{eqCon.17}%
\end{equation}

\textsc{Case 2:} $\lambda=1/2$:
\begin{equation}
X_{1/2}=\left[  \,0,1\,\right]  \text{, and }\mu\text{ \textup{(}}=\mu
_{1/2}\text{\textup{)} is the restriction of Lebesgue measure.}
\label{eqCon.18}%
\end{equation}

\textsc{Case 3:} $\lambda\in\left(  1/2,1\right)  $:
\begin{equation}
X_{\lambda}=\left[  \,0,\frac{\lambda}{1-\lambda}\,\right]  \text{; overlap.}
\label{eqCon.19}%
\end{equation}
In the next section we will study these measures $\mu_{\lambda}$
\textup{(}Case $3$\textup{)} in detail.

In all three cases, the encoding mapping is
\begin{equation}
\pi\left(  \omega_{1}\omega_{2}\dots\right)  =\sum_{i=1}^{\infty}\omega
_{i}\lambda^{i}=\omega_{1}\lambda+\omega_{2}\lambda^{2}+\cdots,\qquad
\omega_{i}\in\mathbb{Z}_{2}=\left\{  0,1\right\}  . \label{eqCon.20}%
\end{equation}
The measure $\mu_{\lambda}$ is determined by the following approximation: If
$E$ is a Borel subset of $X_{\lambda}$, then%
\begin{equation}
\mu_{\lambda}\left(  E\right)  =\lim_{n\rightarrow\infty}2^{-n}\,\#\left\{
\,\left(  \omega_{1}\omega_{2}\dots\omega_{n}\right)  \mid\omega_{1}%
\lambda+\omega_{2}\lambda^{2}+\dots+\omega_{n}\lambda^{n}\in E\,\right\}  .
\label{eqCon.21}%
\end{equation}

\end{example}

\begin{proof}
The only one of the stated conclusions which is not clear from Corollary
\ref{CorCon.EquiMe} is the limit in (\ref{eqCon.21}). But (\ref{eqCon.21}) is
an application of (\ref{eqCon.6}) to the case when the initial measure $\nu$
is $\delta_{0}$ (the Dirac measure), i.e.,
\[
\delta_{0}\left(  E\right)  =\left\{
\begin{array}
[c]{ll}%
1 & \text{\quad if }0\in E,\\
0 & \text{\quad if }0\notin E.
\end{array}
\right.
\]
Note that%
\begin{align*}
T^{n}\delta_{0}  &  =2^{-n}\sum_{\omega_{1}\omega_{2}\dots\omega_{n}}%
\delta_{0}\circ\tau_{\omega_{n}}^{-1}\tau_{\omega_{n-1}}^{-1}\circ\dots
\circ\tau_{\omega_{1}}^{-1}\\
&  =2^{-n}\sum_{\omega_{1}\omega_{2}\dots\omega_{n}}\delta_{\omega_{1}%
\lambda+\omega_{2}\lambda^{2}+\dots+\omega_{n}\lambda^{n}}\,.
\end{align*}
The summation is over $\underbrace{\left\{  0,1\right\}  \times\dots
\times\left\{  0,1\right\}  }_{n\text{ times}}$ for each $n=1,2,\dots$.
\end{proof}

\begin{remark}
\label{RemCon.SumOve}The three cases in \textup{(\ref{eqCon.17}%
)--(\ref{eqCon.19})} are different in one important respect, \emph{overlap};
and we turn to this in the next section.

Summary: \textsc{Overlap:}%
\begin{align*}
&  \text{\textsc{Case 1: }}\lambda\in\left(  0,1/2\right)  : &  &  \tau
_{0}\left(  X_{\lambda}\right)  \cap\tau_{1}\left(  X_{\lambda}\right)
=\varnothing.\vphantom{\left\{ \frac{1}{2}\right\} }\\
&  \text{\textsc{Case 2: }}\lambda=1/2: &  &  \tau_{0}\left(  \left[
\,0,1\,\right]  \right)  \cap\tau_{1}\left(  \left[  \,0,1\,\right]  \right)
=\left\{  \frac{1}{2}\right\}  .\\
&  \text{\textsc{Case 3: }}\lambda\in\left(  1/2,1\right)  : &  &  \tau
_{0}\left(  X_{\lambda}\right)  \cap\tau_{1}\left(  X_{\lambda}\right)
=\left[  \,\lambda,\frac{\lambda^{2}}{1-\lambda}\right]  .
\end{align*}

\end{remark}

\begin{definition}
\label{DefCon.EssOve}Let $\left(  X,\mathcal{B},\mu\right)  $ be a finite
measure space, and let $\left(  \tau_{i}\right)  _{i=1}^{N}$ be a finite
system of measurable endomorphisms, $\tau_{i}\colon X\rightarrow X$,
$i=1,\dots,N$; and suppose $\mu$ is some normalized equilibrium measure. We
then say that the system has \emph{essential overlap} if%
\begin{equation}
\mathop{\sum\sum}_{i\neq j}\mu\left(  \tau_{i}\left(  X\right)  \cap\tau
_{j}\left(  X\right)  \right)  >0. \label{eqCon.22}%
\end{equation}

\end{definition}

In Section \ref{Gen}, we will prove the following:

\begin{theorem}
\label{ThmCon.EssOve}Let $\left(  X,\mathcal{B},\mu\right)  $ and $\left(
\tau_{i}\right)  _{i=1}^{N}$ be as in Definition \textup{\ref{DefCon.EssOve}};
in particular we assume that $\mu$ is some $\left(  \tau_{i}\right)
$-equilibrium measure. We assume further that each $\tau_{i}$ is of finite type.

Let $\mathcal{H}=L^{2}\left(  \mu\right)  $,
\[
F_{i}\colon f\longmapsto\frac{1}{\sqrt{N}}f\circ\tau_{i},
\]
and let $\mathbb{F}=\left(  F_{i}\right)  $ be the corresponding column isometry

Then $\mathbb{F}$ maps onto $\bigoplus_{1}^{N}L^{2}\left(  \mu\right)  $ if
and only if $\left(  \tau_{i}\right)  $ has zero $\mu$-essential overlap.
\end{theorem}

\begin{proof} To better understand the geometric significance of this theorem,
before giving the proof, we work out a particular family of examples
(Sections \ref{Ove} and \ref{Sie} below) in 1D and in 2D.

          The intricate geometric features of IFSs can be understood
nicely by specializing the particular affine transformations making up
the IFS to have a single scale number (which we call $\lambda$). In
the 1D examples, we will have two affine transformations
(\ref{eqOve.1}), and in our 2D examples, three
(\ref{eqSie.3}). (These special 1D examples also go under the name
infinite Bernoulli convolutions. Here we will use them primarily for
illustrating the operator theory behind Theorem \ref{ThmCon.EssOve}.)

         As shown in Section \ref{Sie} below, in passing from 1D to
2D, the possible geometries of the IFS-recursions increase; for
example, new fractions and new gaps may appear simultaneously at each
iteration step. Specifically, (i) fractal (i.e., repeated gaps) and
(ii) ``essential overlap'' co-exist in the 2D examples. Here
"essential overlap" (Definition \ref{DefCon.EssOve}) is of course
defined in terms of the Hutchinson measure $\mu$ ($= \mu_{\lambda}$).

      Our examples will be followed by a proof of Theorem
\ref{ThmCon.EssOve} in section \ref{Gen}. In this concluding
section we further show (Theorem \ref{ThmGen.IsoExt}) that every
IFS with essential overlap (Definition \ref{DefCon.EssOve}) has a
canonical and minimal dilation to one with non-overlap.
\end{proof}

\begin{remark}
\label{RemCon.RepCun}Note that the conclusion of the theorem states that the
operators $S_{i}:=F_{i}^{\ast}$ define a representation of the Cuntz $C^{\ast
}$-algebra $\mathcal{O}_{N}$ if and only if the system has non-essential
overlap, i.e., if and only if $\mu\left(  \tau_{i}\left(  X\right)  \cap
\tau_{j}\left(  X\right)  \right)  =0$ for all $i\neq j$.
\end{remark}

\section{\label{Ove}Operator theory of essential overlap}

The one-dimensional example in the previous section (Example
\ref{ExaCon.xplus1}) helps to clarify the notion of essential overlap. Recall
that in this example, $\lambda$ is chosen such that $1/2<\lambda<1$, and the
two endomorphisms are%
\begin{equation}
\left\{
\begin{aligned} \tau_{0}\left( x\right) &=\lambda x,\\ \tau_{1}\left( x\right) &=\lambda\left( x+1\right) \end{aligned}\right\}
\text{\quad and\quad}\mu=\frac{1}{2}\left(  \mu\circ\tau_{0}^{-1}+\mu\circ
\tau_{1}^{-1}\right)  \label{eqOve.1}%
\end{equation}
with attractor $X_{\lambda}=\operatorname*{supp}\left(  \mu\right)  =\left[
\,0,\lambda/\left(  1-\lambda\right)  \,\right]  $. But the probability
measure $\mu$ ($=\mu_{\lambda}$) on this interval is not the restriction of
Lebesgue measure on $\mathbb{R}$. In fact, $\mu_{\lambda}$ is difficult to
compute explicitly and there appears not to be a closed formula for
$\mu_{\lambda}\left(  \left[  \,\lambda,\lambda^{2}/\left(  1-\lambda\right)
\,\right]  \right)  $, although the Lebesgue measure of the overlap is
$\frac{\lambda\left(  2\lambda-1\right)  }{1-\lambda}$. It is known, for
example, that $\mu_{\lambda}$ is Lebesgue-absolutely continuous for
$\mathrm{a.a.}\,\lambda$; see \cite{Sol95}.

{}From Example \ref{ExaCon.xplus1} it also follows that $\mu_{\lambda}$ is the
distribution of the random variable $\pi_{\lambda}\left(  \omega\right)
=\sum_{i=1}^{\infty}\omega_{i}\lambda^{i}$, $\omega_{i}\in\left\{
0,1\right\}  $, or $\omega=\left(  \omega_{1}\omega_{2}\dots\right)  \in
\Omega=\left\{  0,1\right\}  ^{\mathbb{N}}$. Specifically,
\begin{equation}
P_{1/2}\left(  \left\{  \,\omega\mid\pi_{\lambda}\left(  \omega\right)  \leq
x\,\right\}  \right)  =\int_{-\infty}^{x}d\mu_{\lambda}\left(  y\right)
=\mu_{\lambda}\left(  \left[  \,0,x\,\right]  \right)  =:F_{\lambda}\left(
x\right)  . \label{eqOve.2}%
\end{equation}
Recall from Section \ref{Mul} that $\operatorname*{supp}\left(  \mu_{\lambda
}\right)  =X_{\lambda}=\left[  \,0,\lambda/\left(  1-\lambda\right)
\,\right]  .$

While explicit analytic properties (like bounded variation) satisfied by the
cumulative distribution functions $F_{\lambda}$ (for specific values of
$\lambda$) aren't well understood, some geometric properties may be generated
from the following proposition. It gives an intrinsic scaling identity
(\ref{eqOveNov26.pound}) for $F_{\lambda}$, and a recursive algorithm
(\ref{eqOveNov26.poundpound}) for its computation.

First note that by (\ref{eqOve.2}), for every $\lambda$, the function $x \to
F(x) = F_{\lambda}(x)$ is monotone nondecreasing.

Moreover since the Hutchinson measure $\mu_{\lambda}$ is supported in the
interval $X_{\lambda}$, $F_{\lambda}$ is constantly $0$ on the negative real
line, and it is $1$ on the infinite half line to the right of the bounded
interval $[\,0, \lambda/(1 - \lambda)\,]$.

Of course, $F$ need not be strictly increasing in the interval. But since it
is monotone, by Lebesgue's theorem, it is differentiable almost everywhere
(for a.e.\ $x$, taken with respect to Lebesgue measure). However, a deep
theorem \cite{Sol95} states that, as a measure, $\mu_{\lambda}$ has $L^{2}$
density for a.e.\ $\lambda$ in the open interval $(1/2,1)$; so in particular
for these values of $\lambda$, the Stieltjes measure $\mu_{\lambda
}=dF_{\lambda}$ is relatively absolutely continuous with respect to Lebesgue measure.

\subsection{Symmetry}\label{Ove.Sym}

The next result (needed later!) shows that with the choice of equal weights
$\frac{1}{2}$--$\frac{1}{2}$ in (\ref{eqOve.1}), the random variable
$\pi_{\lambda}\colon\Omega\rightarrow\left[  \,0,\lambda/\left(
1-\lambda\right)  \,\right]  $ from (\ref{eqCon.20}) is symmetric around the
midpoint%
\begin{equation}
\frac{\lambda}{2\left(  1-\lambda\right)  }=\int_{0}^{\frac{\lambda}%
{1-\lambda}}x\,d\mu_{\lambda}\left(  x\right)  . \label{eqOveNov29.new1}%
\end{equation}
The symmetry property which is made precise in the next lemma reflects that
the $x$-values are corners of a hypercube: Repeated folding with shrinking
intervals each with excess length.

\begin{lemma}
[Symmetry]\label{LemOve.Symmetry}Let $\lambda\in\left(  1/2,1\right)  $, and
let $\mu=\mu_{\lambda}$ and $P_{1/2}$ be as described in
\textup{(\ref{eqOve.1})} and \textup{(\ref{eqOve.2})}. Recall that
$X_{\lambda}=\operatorname*{supp}\left(  \mu_{\lambda}\right)  $ is the closed
interval $\left[  \,0,b\left(  \lambda\right)  \,\right]  $ where $b\left(
\lambda\right)  :=\lambda/\left(  1-\lambda\right)  $.

Then for all $x\leq\frac{1}{2}b\left(  \lambda\right)  $ the two tail-ends of
the distribution are the same, i.e.,%
\begin{equation}
P_{1/2}\left(  \left\{  \,\omega\mid\pi_{\lambda}\left(  \omega\right)
<x\,\right\}  \right)  =P_{1/2}\left(  \left\{  \,\omega\mid\pi_{\lambda
}\left(  \omega\right)  >b\left(  \lambda\right)  -x\,\right\}  \right)  .
\label{eqOveNov29.new3}%
\end{equation}

\end{lemma}

\begin{proof}
We will prove (\ref{eqOveNov29.new3}) in the equivalent form%
\begin{equation}
P_{1/2}\left(  \left\{  \,\omega\mid\pi_{\lambda}\left(  \omega\right)  \geq
x\,\right\}  \right)  =P_{1/2}\left(  \left\{  \,\omega\mid\pi_{\lambda
}\left(  \omega\right)  \leq b\left(  \lambda\right)  -x\,\right\}  \right)  .
\label{eqOveNov29.new4}%
\end{equation}
The two events we specify on the left-hand side and the right-hand side in
(\ref{eqOveNov29.new4}) are described by the respective conditions%
\begin{equation}
\sum_{i=1}^{\infty}\omega_{i}\lambda^{i}\geq x \label{eqOveNov29.new5}%
\end{equation}
and%
\begin{equation}
\sum_{i=1}^{\infty}\left(  1-\omega_{i}\right)  \lambda^{i}\geq x.
\label{eqOveNov29.new6}%
\end{equation}
But since the \textquotedblleft fair-coin\textquotedblright\ Bernoulli measure
$P_{1/2}$ on $\Omega$ was chosen, the two sequences of independent random
variables
\[
\omega_{1},\;\omega_{2},\;\omega_{3},\;\dots
\]
and%
\[
1-\omega_{1},\;1-\omega_{2},\;1-\omega_{3},\;\dots
\]
are equi-distributed. Hence the numbers on the two sides in
(\ref{eqOveNov29.new4}) are the same.

The formula (\ref{eqOveNov29.new1}) for the mean follows directly from
(\ref{eqOve.1}) as follows: Set $M_{1}=M_{1}\left(  \lambda\right)  =\int
x\,d\mu_{\lambda}\left(  x\right)  $. Then by (\ref{eqOve.1}),%
\begin{align*}
M_{1}  &  =\frac{1}{2}\left(  \int\left(  \lambda x\right)  \,d\mu_{\lambda
}\left(  x\right)  +\int\lambda\left(  x+1\right)  \,d\mu_{\lambda}\left(
x\right)  \right) \\
&  =\lambda M_{1}+\frac{\lambda}{2}.
\end{align*}
Hence $M_{1}=\frac{\lambda}{2\left(  1-\lambda\right)  }=\frac{1}{2}b\left(
\lambda\right)  $ as claimed.
\end{proof}

\begin{remark}
\label{RemOve.CascadeMid}
Actually every ``cascade approximant'' \textup{(}see Proposition
\textup{\ref{ProOveNov26.1}} below\/\textup{)} is symmetric in a similar way around its
midpoint $\left(  1/2\right)  \left(  \lambda+ \lambda^{2} + \dots+
\lambda^{n}\right)  $.
\end{remark}

\begin{remark}
\label{RemOve.ManyMoments}The argument at the end of the proof of the lemma
extends to yield a formula for all the moments
\[
M_{n}\left(  \lambda\right)  =\int x^{n}\,d\mu_{\lambda}\left(  x\right)  ,
\]
beginning with%
\[
M_{2}\left(  \lambda\right)  =\frac{\lambda^{2}}{2\left(  1-\lambda\right)
^{2}\cdot\left(  1+\lambda\right)  }= \frac{2 M_{1}\left(  \lambda\right)
^{2} }{\left(  1 + \lambda\right)  }
\]
and%
\[
M_{3}\left(  \lambda\right)  =\frac{\lambda^{3}\cdot\left(  \lambda
+2\cdot\left(  1-\lambda^{2}+\lambda^{3}\right)  \right)  }{4\cdot\left(
1-\lambda^{3}\right)  \left(  1-\lambda^{2}\right)  \left(  1-\lambda\right)
}.
\]
The formula for $n>1$ is recursive, and can be worked out from the binomial distribution.

There is a considerable amount of recent work \textup{(}see, e.g.,
\cite{CuFi05}\textup{)} on moments in operator theory, and it would be
interesting to explore the operator-theoretic significance of our present
\textquotedblleft overlap-moments.\textquotedblright
\end{remark}

\begin{remark}
\label{RemOve.AntiAtom}By the argument of the previous remark, the
moment-generating function
\[
\hat{\mu}_{\lambda}\left(  t\right)  :=\int e^{itx}\,d\mu_{\lambda}\left(
x\right)
\]
can be shown to have the following infinite-product expansion:%
\[
\hat{\mu}_{\lambda}\left(  t\right)  =e^{itM_{1}\left(  \lambda\right)  }%
\prod_{n=1}^{\infty}\cos\left(  \frac{t\lambda^{n}}{2}\right)  .
\]
Using Wiener's test on this, it follows that none of the measures
$\mu_{\lambda}$ have atoms, i.e., that $\mu_{\lambda}\left(  \left\{
x\right\}  \right)  =0$ for all $x$, and for all $\lambda\in\left[
1/2,1\right)  $.
\end{remark}

\begin{remark}
[Figure generation, hypercubes, and symmetry, by Brian Treadway]%
\label{RemOve.Hypercube}The $x$-values in the figures where steps occur in $F$
are generated using a recursive \textquotedblleft outer
sum\textquotedblright\ construction:%
\begin{multline*}
\{0\}\rightarrow\{\{0\},\{\lambda\}\}\rightarrow\{\{\{0\},\{\lambda
^{2}\}\},\{\{\lambda\},\{\lambda+\lambda^{2}\}\}\}\\
{}\rightarrow\{\{\{\{0\},\{\lambda^{3}\}\},\{\{\lambda^{2}\},\{\lambda
^{2}+\lambda^{3}\}\}\},\{\{\{\lambda\},\{\lambda+\lambda^{3}\}\},\{\{\lambda
+\lambda^{2}\},\{\lambda+\lambda^{2}+\lambda^{3}\}\}\}\}\\
{}\rightarrow\cdots.
\end{multline*}
This nested list of the $2^{n}$ sums has the structure of the coding map---in
fact, if it is expressed with \textquotedblleft tensor
indices\textquotedblright\ instead of nested braces, the indices are just our
$\omega$'s. The hypercube idea comes from thinking of each $\omega_{i}$ as a
coordinate in an orthogonal direction. Opposite corners then have $x$-values
that sum to $(\lambda+\lambda^{2}+\dots+\lambda^{n})$, so the two
tails may be matched point by point. Hence the $x$-values are corners of a hypercube, and
hence symmetry.

To make the plots of $F\left(x\right)$, the nested list above is 
``flattened'' to a single list and sorted numerically.  The \textit{Mathematica} program that
generates the sorted list is
\begin{center}
\texttt{Sort[Flatten[Nest[Outer[Plus, \{0, l\}, l \#]\&, \{0\}, n]], OrderedQ[\{N[\#1], N[\#2]\}]\&]},
\end{center}
where ``\texttt{l}'' stands for $\lambda$ and ``\texttt{n}'' is the level
of iteration in the cascade algorithm.
\end{remark}

Aside from Erd\H{o}s's theorem \cite{Erd40} about the case of $\lambda$ equal
to the golden ratio, where $\mu_{\lambda}$ is singular, precious little is
known about $\mu_{\lambda}$ for other specific $\lambda$ values, for example
when $\lambda$ is rational.

Two interesting values of $\lambda$ are $\lambda= (\sqrt{5} - 1)/2$
(Fig.\ \ref{FigOve.Flambda}) and $\lambda= 3/4$
(Fig.\ \ref{Figfplotabcdefg}). The cases are qualitatively different both
with regard to absolute continuity and overlap and with respect to Fourier
bases; see \cite{DuJo06e}.

In the known results \cite{DuJo06e} for which $L^{2}(\mu_{\lambda})$ has a
Fourier orthonormal basis (ONB), $\lambda$ is rational. By a Fourier basis, we
mean an ONB in $L^{2}(\mu_{\lambda})$ consisting of complex exponentials.

Because of \cite{Erd40}, $\lambda= (\sqrt{5} - 1)/2$ is likely to have its
$F_{\lambda}$ a little less ``differentiable'' than the $F_{\lambda}$ for $\lambda= 3/4$.

The following details will work as an iterative and cascading algorithm for $F
= F_{\lambda}$ when $\lambda$ is fixed. Our algorithm is initialized so as to
take advantage of (\ref{eqCon.21}) above. It is illustrated in the figures;
see especially Figure \ref{FigOve.FlambdaCascade}.

\begin{proposition}
\label{ProOveNov26.1} Let $\lambda$ be given in the open interval $(1/2, 1)$,
and set $F = F_{\lambda}$ as a function on $\mathbb{R}$. Conclusions:\renewcommand{\theenumi}{\alph{enumi}}

\begin{enumerate}
\item \label{ProOveNov26.1(a)} Then the function $x \to F(x)$ in
\textup{(\ref{eqOve.2})} satisfies
\begin{equation}
F(x) = \frac12\left(  F\left(  \frac{x}{\lambda}\right)  + F\left(  \frac{x -
\lambda}{\lambda}\right)  \right)  ; \label{eqOveNov26.pound}%
\end{equation}
or equivalently, for expansive scaling number $s =1/\lambda$,
\[
F(x) = \frac12\left(  F\left(  s x\right)  + F\left(  s x - 1\right)  \right)
.
\]

\item \label{ProOveNov26.1(b)} Then the following iterative and cascading
algorithm holds: Initializing $F_{0}$ by setting $F_{0}$ to be the Heaviside
function
\begin{align*}
F_{0}\left(  x\right)   &  =0\text{\qquad for }x<0\text{, and}\\
F_{0}\left(  x\right)   &  =1\text{\qquad for }x\geq0,
\end{align*}
we get the following recursion: $F_{0}$, $F_{1}$, $\dots$, etc., with
\begin{equation}
F_{n+1}\left(  x\right)  =\frac{1}{2}\left(  F_{n}\left(  sx\right)
+F_{n}\left(  sx-1\right)  \right)  ,\qquad n=0,1,2,\dots,
\label{eqOveNov26.poundpound}%
\end{equation}
or equivalently
\[
F_{n+1}\left(  x\right)  =\frac{1}{2}\left(  F_{n}\left(  \frac{x}{\lambda
}\right)  +F_{n}\left(  \frac{x-\lambda}{\lambda}\right)  \right)  .
\]

Moreover, $F_{n}\left(  x\right)  =1$ holds for $x>\lambda/(1-\lambda)$, and
for all $n$; and the sequence $F_{n}$ is convergent pointwise in the closed
interval
\begin{equation}
X=X_{\lambda}=\left[  \,0,\frac{\lambda}{1-\lambda}\,\right]  =\left[
\,0,\frac{1}{s-1}\,\right]  . \label{eqOveNov26.interval}%
\end{equation}
For each $x$, this convergence is monotone, and
\[
F\left(  x\right)  =\inf_{n}F_{n}\left(  x\right)  .
\]

\end{enumerate}
\end{proposition}

\begin{proof}
The scaling identity (\ref{eqOveNov26.pound}) follows from the following facts
(see \cite{Hut81}): The limit formula (\ref{eqCon.6}); the fixed-point
property (\ref{eqCon.3}), i.e., $T(\mu)=\mu$; and the fact that the
equilibrium measure $\mu$ is supported in the closed interval
\[
X=X_{\lambda}=\left[  \,0,\frac{\lambda}{1-\lambda}\,\right]  .
\]
See also formula (\ref{eqCon.21}).

\emph{Ad} (\ref{ProOveNov26.1(a)}): Specifically, for $x\in X_{\lambda}$, we
have\settowidth{\equallabelwidth}{$\scriptstyle\,\operatorname*{supp}\left(  \mu\right)  \subset X_{\lambda}\,$}
\begin{align*}
F\left(  x\right)   &  \underset
{\makebox[\equallabelwidth]{$\scriptstyle\text{by (\ref{eqOve.2})}$}}{=}%
\int_{X_{\lambda}}\chi_{\left[  \,0,x\,\right]  }\left(  y\right)
\,d\mu\left(  y\right) \\
&  \underset
{\makebox[\equallabelwidth]{$\scriptstyle\text{by (\ref{eqOve.1})}$}}{=}%
\frac{1}{2}\left(  \int_{0}^{\frac{\lambda}{1-\lambda}}\chi_{\left[
\,0,x\,\right]  }\left(  \tau_{0}\left(  y\right)  \right)  \,d\mu\left(
y\right)  +\int_{0}^{\frac{\lambda}{1-\lambda}}\chi_{\left[  \,0,x\,\right]
}\left(  \tau_{1}\left(  y\right)  \right)  \,d\mu\left(  y\right)  \right) \\
&  \underset{\,\operatorname*{supp}\left(  \mu\right)  \subset X_{\lambda}%
\,}{=}\frac{1}{2}\left(  \int_{0}^{\frac{x}{\lambda}}d\mu\left(  y\right)
+\int_{0}^{\frac{x-\lambda}{\lambda}}d\mu\left(  y\right)  \right) \\
&  \underset
{\makebox[\equallabelwidth]{$\scriptstyle\text{by (\ref{eqOve.2})}$}}{=}%
\frac{1}{2}\left(  F\left(  \frac{x}{\lambda}\right)  +F\left(  \frac
{x-\lambda}{\lambda}\right)  \right)  .
\end{align*}
For the evaluation of the second integral, note: $\tau_{1}\left(  y\right)
\in\left[  \,0,x\,\right]  \iff-1\leq y\leq\frac{x-\lambda}{\lambda}$. But
since $\operatorname*{supp}\left(  \mu\right)  \subset X_{\lambda}$, we get
that%
\[
\int_{-1}^{\frac{x-\lambda}{\lambda}}d\mu\left(  y\right)  =\int_{0}%
^{\frac{x-\lambda}{\lambda}}d\mu\left(  y\right)  .
\]

\emph{Ad} (\ref{ProOveNov26.1(b)}): Figures \ref{FigOve.FlambdaCascade}(a),
(b), (c), $\dots$ are sketches of the successive
functions $F_{0}$, $F_{1}$, $F_{2}$, $\dots$ in the case of $\lambda=\frac{\sqrt
{5}-1}{2}$, i.e., the reciprocal of the golden
ratio $\phi= \left(  \sqrt{5} + 1\right)  /2$.

As indicated, the
sequence is monotone, i.e., $F_{n+1}\left(  x\right)  \leq F_{n}\left(
x\right)  $ holds; as will be proved.


\begin{figure}
\setlength{\unitlength}{1bp}
\begin{picture}(324,117)(0,-45)
\put(0,0){\includegraphics[bb=88 4 241 76,width=153\unitlength]{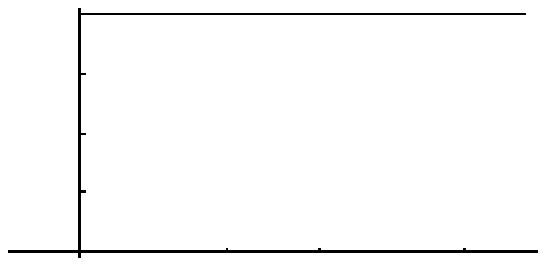}}
\put(18,70.4){\makebox(0,0)[r]{$\scriptstyle1$}}
\put(18,36){\makebox(0,0)[r]{$\scriptstyle1/2$}}
\put(18,0.5){\makebox(0,0)[tr]{$\scriptstyle0$}}
\put(63.5,0.5){\makebox(0,0)[t]{$\scriptstyle\lambda$}}
\put(90,0.5){\makebox(0,0)[t]{$\scriptstyle1$}}
\put(132,0.5){\makebox(0,0)[t]{$\scriptstyle b$}}
\put(76,-13){\makebox(0,0)[t]{(a). $n=0$}}
\put(171,0){\includegraphics[bb=88 4 241 76,width=153\unitlength]{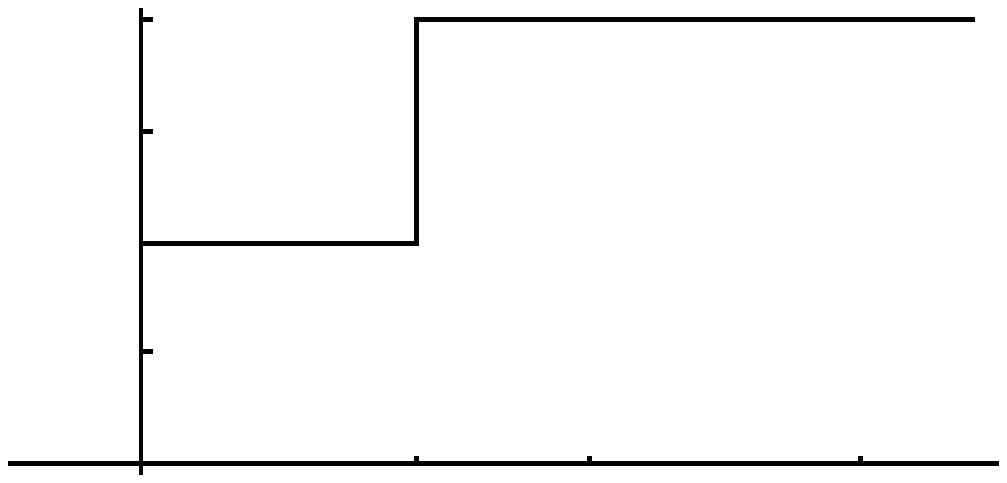}}
\put(189,70.4){\makebox(0,0)[r]{$\scriptstyle1$}}
\put(189,36){\makebox(0,0)[r]{$\scriptstyle1/2$}}
\put(189,0.5){\makebox(0,0)[tr]{$\scriptstyle0$}}
\put(234.5,0.5){\makebox(0,0)[t]{$\scriptstyle\lambda$}}
\put(261,0.5){\makebox(0,0)[t]{$\scriptstyle1$}}
\put(303,0.5){\makebox(0,0)[t]{$\scriptstyle b$}}
\put(247,-13){\makebox(0,0)[t]{(b). $n=1$}}
\end{picture}
\begin{picture}(324,117)(0,-45)
\put(0,0){\includegraphics[bb=88 4 241 76,width=153\unitlength]{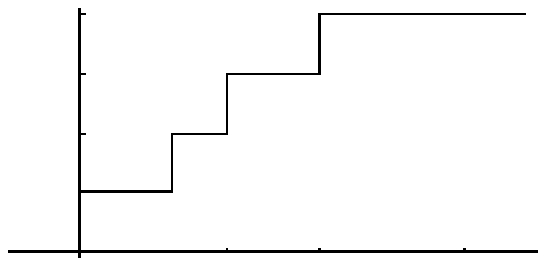}}
\put(18,70.4){\makebox(0,0)[r]{$\scriptstyle1$}}
\put(18,36){\makebox(0,0)[r]{$\scriptstyle1/2$}}
\put(18,0.5){\makebox(0,0)[tr]{$\scriptstyle0$}}
\put(63.5,0.5){\makebox(0,0)[t]{$\scriptstyle\lambda$}}
\put(90,0.5){\makebox(0,0)[t]{$\scriptstyle1$}}
\put(132,0.5){\makebox(0,0)[t]{$\scriptstyle b$}}
\put(76,-13){\makebox(0,0)[t]{(c). $n=2$}}
\put(171,0){\includegraphics[bb=88 4 241 76,width=153\unitlength]{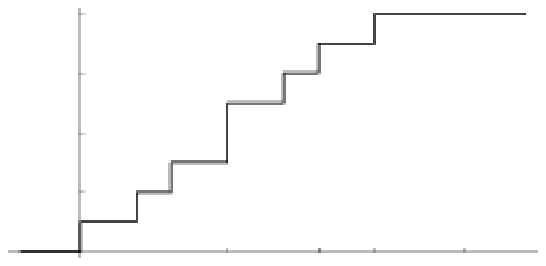}}
\put(189,70.4){\makebox(0,0)[r]{$\scriptstyle1$}}
\put(189,36){\makebox(0,0)[r]{$\scriptstyle1/2$}}
\put(189,0.5){\makebox(0,0)[tr]{$\scriptstyle0$}}
\put(234.5,0.5){\makebox(0,0)[t]{$\scriptstyle\lambda$}}
\put(261,0.5){\makebox(0,0)[t]{$\scriptstyle1$}}
\put(277.5,0.5){\makebox(0,0)[t]{$\scriptstyle b'$}}
\put(303,0.5){\makebox(0,0)[t]{$\scriptstyle b$}}
\put(247,-13){\makebox(0,0)[t]{(d). $n=3$}}
\end{picture}
\caption{The series of cascade approximants $F_{n}$, $n=0,1,2,\dots$, to the
cumulative distribution function $F_{\lambda}$ in the case when
$\lambda=\left(  \sqrt{5}-1\right)  /2$.
The marks on the horizontal axis in addition to $\lambda$ and $1$ are
$b'=\lambda+\lambda^{2}+\dots+\lambda^{n}$, the endpoint of the support of the $n$'th cascade iteration $F_{n}$, 
and $b=\lambda/(1-\lambda)$, the endpoint of the support of $F_{\lambda}$.
In the case of $F_{3}$ for this particular ``golden'' value of
$\lambda$, it may be observed that the set of node-points $N_{3}\left(  \lambda\right)
=\left\{  0,\lambda^{3},\lambda^{2},\lambda,\lambda+\lambda^{3},\lambda
+\lambda^{2},2\lambda\right\}  $, and $\#\,N_{3}\left(  \lambda\right)  =7$, not $8$:
one of the steps is doubled, due to the fact that $\lambda^{2}+\lambda^{3}=\lambda$.}
\label{FigOve.FlambdaCascade}
\end{figure}

We now prove the assertion $F_{n}\left(  x\right)  \equiv1$ for all $n$ and
all $x>\lambda/\left(  1-\lambda\right)  =\lambda+\lambda^{2}+\lambda
^{3}+\cdots$. As before, $\lambda\in\left(  1/2,1\right)  $ is fixed, and
$\lambda/\left(  1-\lambda\right)  $ is the right-hand endpoint in the
interval $X_{\lambda}$.

The assertion follows by induction. It holds for $F_{0}$ since $F_{0}$ is the
Heaviside function. Now suppose it holds for $n$. Let $x>\lambda/\left(
1-\lambda\right)  $ be given. Then
\[
\frac{x}{\lambda}>\frac{1}{1-\lambda}>\frac{\lambda}{1-\lambda},
\]
and%
\[
\frac{x-\lambda}{\lambda}>\frac{\frac{\lambda}{1-\lambda}-\lambda}{\lambda
}=\frac{\lambda}{1-\lambda};
\]
so%
\[
F_{n+1}\left(  x\right)  =\frac{1}{2}\left(  F_{n}\left(  \frac{x}{\lambda
}\right)  +F_{n}\left(  \frac{x-\lambda}{\lambda}\right)  \right)  =\frac
{1}{2}\left(  1+1\right)  =1,
\]
completing the induction.

The separate assertion about pointwise convergence%
\begin{equation}
\lim_{n\rightarrow\infty}F_{n}\left(  x\right)  =F\left(  x\right)
\label{eqOveNov26.newpound1}%
\end{equation}
follows from the stronger fact: For each $x$, the sequence $\left(
F_{n}\left(  x\right)  \right)  _{n}$ is monotone, i.e.,%
\begin{equation}
F_{n}\left(  x\right)  \leq F_{n-1}\left(  x\right)  .
\label{eqOveNov26.newpound2}%
\end{equation}
Induction: Clearly (\ref{eqOveNov26.newpound2}) holds if $n=1$. Now suppose it
holds up to $n$. Then by the recursion,
\[
F_{n}\!\left(  x\right)  -F_{n+1}\!\left(  x\right)  =\frac{1}{2}%
\Biggl(\left(  F_{n-1}\!\left(  \frac{x}{\lambda}\right)  -F_{n}\!\left(
\frac{x}{\lambda}\right)  \right)  +\left(  F_{n-1}\!\left(  \frac{x-\lambda
}{\lambda}\right)  -F_{n}\!\left(  \frac{x-\lambda}{\lambda}\right)  \right)
\Biggr).
\]
The conclusion (\ref{eqOveNov26.newpound2}) now follows for $n+1$, and the
proof is complete.
\end{proof}

\subsection{Measuring overlap}\label{measuringoverlap}

\begin{remark}
\label{RemOveNov26.pound1}The three figures
Figs.~\textup{\ref{FigOve.FlambdaCascade}(a),
(b), (c)} taken by themselves offer an
oversimplification in that the ordering of the node-points%
\[
N_{n}\left(  \lambda\right)  :=\left\{  \,\omega_{1}\lambda+\omega_{2}%
\lambda^{2}+\dots+\omega_{n}\lambda^{n}\mid\omega_{i}\in\left\{  0,1\right\}
\,\right\}
\]
is unique only up to $n=2$, i.e.,
\[
0<\lambda^{2}<\lambda<\lambda+\lambda^{2}%
\]
holds for all $\lambda\in\left(  1/2,1\right)  $. But in general for $n>2$,
the ordering of the $2^{n}$ points in $N_{n}\left(  \lambda\right)  $ is
subtle. For example, for $n=3$, we have%
\begin{equation}
\left\{
\begin{aligned} \lambda&<\lambda^{2}+\lambda^{3}&&\text{\quad if }\lambda>\frac{\sqrt{5}-1}{2}, \\ \lambda&=\lambda^{2}+\lambda^{3}&&\text{\quad if }\lambda=\frac{\sqrt{5}-1}{2}, \\ \lambda&>\lambda^{2}+\lambda^{3}&&\text{\quad if }\lambda<\frac{\sqrt{5}-1}{2}, \end{aligned}\right.
\label{eqOveNov26.pound3}%
\end{equation}
indicating that even in $N_{3}$ order-reversal may occur depending on the
chosen value of $\lambda$ in $\left(  1/2,1\right)  $.

The list in \textup{(\ref{eqOveNov26.pound3})} further shows that the points
in each set $N_{n}\left(  \lambda\right)  $ for $n>2$ typically occur with multiplicity.
\end{remark}

The assertion in (\ref{eqOveNov26.pound3}) for the special case $\lambda
=\left(  \sqrt{5}-1\right)  /2$ states that%
\[
\pi\left(  1000\dots\right)  =\pi\left(  011000\dots\right)  ,
\]
where $\pi=\pi_{\lambda}$ is the encoding mapping $\pi_{\lambda}\colon
\Omega\rightarrow X_{\lambda}$ from Lemma \ref{LemCon.CodCon} and
(\ref{eqCon.20}). While it is known that for all $\lambda\in\left(
1/2,1\right)  $, $\pi_{\lambda}$ is $\infty$--$1$, the infinite sets
$\pi_{\lambda}^{-1}\left(  x\right)  $ are not well understood.

\begin{proposition}
\label{ProOveNov28.OneThird}If $\lambda=\left(  \sqrt{5}-1\right)  /2$ and
$\mu=\mu_{\lambda}$ is normalized, i.e., $\mu\left(  X\right)  =1$, then
\[
\mu\left(  \tau_{0}\left(  X\right)  \cap\tau_{1}\left(  X\right)  \right)
=\frac{1}{3}.
\]

\end{proposition}

\begin{proof}
[Proof of Proposition \textup{\ref{ProOveNov28.OneThird}}]{}From
(\ref{eqOve.1}) we see that $\tau_{0}\left(  X\right)  =\left[  \,0,\lambda
^{2}/\left(  1-\lambda\right)  \,\right]  $ and $\tau_{1}\left(  X\right)
=\left[  \,\lambda,\lambda/\left(  1-\lambda\right)  \,\right]  $. A symmetry
consideration (Lemma \ref{LemOve.Symmetry}) further shows that $\mu\left(
\tau_{0}\left(  X\right)  \right)  =\mu\left(  \tau_{1}\left(  X\right)
\right)  $. We will compute $\mu\left(  \tau_{1}\left(  X\right)  \right)  $
using (\ref{eqCon.21}). In fact, we show that%
\begin{equation}
\mu\left(  \tau_{1}\left(  X\right)  \right)  =\frac{2}{3}.
\label{eqOveNov28.poundprime}%
\end{equation}
Since $\mu\left(  \tau_{0}\left(  X\right)  \cup\tau_{1}\left(  X\right)
\right)  =\mu\left(  X\right)  =1$, and $1=\mu\left(  \text{union}\right)
=2\mu\left(  \tau_{1}\left(  X\right)  \right)  -\mu\left(  \text{overlap}%
\right)  $, we get%
\[
\mu\left(  \tau_{0}\left(  X\right)  \cap\tau_{1}\left(  X\right)  \right)
=\frac{4}{3}-1=\frac{1}{3},
\]
as claimed.

Set $b=b_{\lambda}=\lambda/\left(  1-\lambda\right)  $. Since $\lambda
^{2}+\lambda-1=0$, we then get $b=1/\lambda=2/\left(  \sqrt{5}-1\right)  $,
and therefore $\tau_{1}\left(  X\right)  =\left[  \,\lambda,b\,\right]  $.

We now turn to (\ref{eqOveNov28.poundprime}). By (\ref{eqCon.21}),
\[
\mu\left(  \tau_{1}\left(  X\right)  \right)  =\lim_{n\rightarrow\infty
}\left(  2^{-n}\right)  \cdot\#\left(  \pi^{-1}\left(  N_{n}\left(
\lambda\right)  \cap\left[  \,\lambda,b\,\right]  \right)  \right)  .
\]
Recall that%
\[
N_{n}\left(  \lambda\right)  =\left\{  \,\sum_{i=1}^{n}\omega_{i}\lambda
^{i}\biggm|\omega=\left(  \omega_{1}\dots\omega_{n}\right)  \in\left\{
0,1\right\}  ^{n}\,\right\}
\]
for all $n\in\mathbb{N}$. We claim that for $n>2$,
\begin{multline}
\pi^{-1}\left(  N_{n}\left(  \lambda\right)  \cap\left[  \,\lambda,b\,\right]
\right)  =\left\{  \,\omega\in\left\{  0,1\right\}  ^{n}\mid\omega
_{1}=1\,\right\} \\
{}\cup\left\{  \,\omega\in\left\{  0,1\right\}  ^{n}\mid\omega_{1}%
=0,\;\omega_{2}=\omega_{3}=1\,\right\}  \cup\cdots,
\label{eqOveNov28.poundprimeprime}%
\end{multline}
where the union of the individual sets on the right-hand side is clearly
disjoint. The sets indicated by \textquotedblleft$\cup\cdots$%
\textquotedblright\ on the right-hand side in
(\ref{eqOveNov28.poundprimeprime}) have the form%
\[
\bigcup_{k}\left\{  \,\omega\mid\omega_{2i-1}=0,\;\omega_{2i}=1,\;i\leq
k\text{; and }\omega_{2k+1}=1\right\}  .
\]
The contribution to these sets is $\geq\lambda^{2}+\lambda^{4}+\lambda
^{6}+\dots+\lambda^{2k}+\lambda^{2k+1}=\lambda$, where we used that
$\lambda+\lambda^{2}=1$. It follows that%
\[
\lim_{n\rightarrow\infty}\left(  2^{-n}\right)  \cdot\#\left(  \pi^{-1}\left(
N_{n}\left(  \lambda\right)  \cap\left[  \,\lambda,b\,\right]  \right)
\right)  =2^{-1}+2^{-3}+2^{-5}+2^{-7}+\cdots=\frac{2}{3}%
\]
as claimed.

It is clear that the right-hand side in (\ref{eqOveNov28.poundprimeprime}) is
contained in $\pi^{-1}\left(  N_{n}\left(  \lambda\right)  \cap\left[
\,\lambda,b\,\right]  \right)  $ for sufficiently large $n>2$. The assertion
that they are equal follows from the fact that%
\begin{equation}
\lambda^{3}+\lambda^{4}+\dots+\lambda^{m}<\lambda
\label{eqOveNov28.poundprimeprimeprime}%
\end{equation}
for all $m>3$. But using $1-\lambda=\lambda^{2}$, note that
(\ref{eqOveNov28.poundprimeprimeprime}) is equivalent to $1-\lambda^{m-2}<1$,
which is clearly true. This proves the proposition.

Recall that the overlap in $\mu_{\lambda}$-measure is%
\[
\mu_{\lambda}\left(  \left[  \,\lambda,b\left(  \lambda\right)  -\lambda
\,\right]  \right)  =P_{1/2}\left(  \left\{  \,\omega\in\Omega\mid\pi
_{\lambda}\left(  \omega\right)  \in\left[  \,\lambda,b\left(  \lambda\right)
-\lambda\,\right]  \,\right\}  \right)  .
\]
This means that the contributions to $\pi_{\lambda}^{-1}\left(  \left[
\,\lambda,b\left(  \lambda\right)  -\lambda\,\right]  \right)  $ with
$P_{1/2}$-measure equal to zero may be omitted in the accounting
(\ref{eqOveNov28.poundprimeprime}) above.

However, even for $\lambda=\left(  \sqrt{5}-1\right)  /2$, even the infinite
set $\pi_{\lambda}^{-1}\left(  \left\{  \lambda\right\}  \right)  $ has
interesting dynamics. But its contribution to the overlap in $\mu$-measure is
\[
\mu_{\lambda}\left(  \left\{  \lambda\right\}  \right)  =0;
\]
see also Remark \ref{RemOve.AntiAtom}.\medskip

\noindent\textbf{Notation.} Set%
\begin{align*}
w  &  :=\left(  \,0\,1\,\right)  ,\\
\underline{0}  &  :=(\underbrace{0\,0\,0\,\dots}_{\infty\text{ repetition}%
})\text{, and}\\
\underline{1}  &  :=(\underbrace{1\,1\,1\,\dots}_{\infty\text{ repetition}}).
\end{align*}
Then $\pi_{\lambda}^{-1}\left(  \left\{  \lambda\right\}  \right)  $ contains
the following infinite lists:%
\begin{align*}
&  \left(  \,1\,\underline{0}\,\right)  ,\\
&  \left(  \,w\,1\,\underline{0}\,\right)  ,\\
&  \left(  \,w\,w\,1\,\underline{0}\,\right)  ,\\
&  \left(  \,w\,w\,w\,1\,\underline{0}\,\right)  ,\\
&  \vdots,
\end{align*}
etc., and%
\begin{align*}
&  \left(  \,0\,0\,\underline{1}\,\right)  ,\\
&  \left(  \,w\,0\,0\,\underline{1}\,\right)  ,\\
&  \left(  \,w\,w\,0\,0\,\underline{1}\,\right)  ,\\
&  \left(  \,w\,w\,w\,0\,0\,\underline{1}\,\right)  ,\\
&  \vdots,
\end{align*}
etc.
\end{proof}

\begin{remark}
\label{RemOveDec1.SiVe} The fact that for $\lambda= \left(  \sqrt{5}
-1\right)  /2$ ($= \phi^{-1} $, $\phi= {}$golden ratio) the overlap measured
in the Hutchinson measure is $1/3$ appears also to follow from \cite{SiVe98}.
In \cite[Cor.~1.2, p.~220]{SiVe98}, entirely about the golden shift, Sidorov
and Vershik do get $1/3$ by a completely different argument: they introduce a
transition matrix on a combinatorial tree, and when translated into our
$\lambda= \left(  \sqrt{5} -1\right)  /2$ example, their Sidorov--Vershik tree
is then Fibonacci. That is key to their computations.

In contrast, our method is general and applies to general metric spaces: IFSs
with overlap. Even when specialized to 1D, for the special case of
(\ref{eqOve.1}), our method has the advantage (see Corollary
\ref{CorOveNov29.LowerBound} and Remark \ref{RemOve.GoldenBound} below) of
estimating the overlap in Hutchinson measure also when $\lambda$ is not
$\phi^{-1 }$, i.e., is not ``golden.''
\end{remark}

\begin{remarks}
\label{RemOve.MeasureForMeasure}%
\setcounter{enumi}{0}\renewcommand{\theenumi}{\alph{enumi}}\refstepcounter
{enumi}\label{RemOve.MeasureForMeasure(1)}%
\textup{(\ref{RemOve.MeasureForMeasure(1)}) The Lebesgue }measure of the
intersection $\tau_{0}\left(  X\right)  \cap\tau_{1}\left(  X\right)  $ is
$\frac{\lambda^{2}}{1-\lambda}-\lambda$, which for $\lambda=\frac{\sqrt{5}%
-1}{2}$ works out to%
\[
\operatorname{Leb}\left(  \tau_{0}\left(  X\right)  \cap\tau_{1}\left(
X\right)  \right)  =\frac{3-\sqrt{5}}{2}.
\]

\refstepcounter{enumi}\label{RemOve.MeasureForMeasure(2)}%
\textup{(\ref{RemOve.MeasureForMeasure(2)}) Since }%
\[
\frac{3-\sqrt{5}}{2}>\frac{1}{3},
\]
we conclude from the proposition that the Hutchinson measure of the
intersection is the smaller of the two.

\refstepcounter{enumi}\label{RemOve.MeasureForMeasure(3)}%
\textup{(\ref{RemOve.MeasureForMeasure(3)}) The argument from the proof of the
proposition extends to the IFS }$\tau_{k}^{\left(  \lambda\right)  }\left(
x\right)  :=\lambda\left(  x+k\right)  $, $k\in\left\{  0,1\right\}  $, for
all $\lambda\in\left(  1/2,1\right)  $, and we conclude that there is
essential overlap for all $\lambda$ in $\left(  1/2,1\right)  $, but an
explicit formula for $\mu_{\lambda}\left(  \tau_{0}^{\left(  \lambda\right)
}\left(  X_{\lambda}\right)  \cap\tau_{1}^{\left(  \lambda\right)  }\left(
X_{\lambda}\right)  \right)  $ ($>0$) is not known.
\end{remarks}

The following is a consequence of the argument in the proof of Proposition
\ref{ProOveNov28.OneThird}.

\begin{corollary}
\label{CorOveNov29.LowerBound}%
\setcounter{enumi}{0}\renewcommand{\theenumi}{\alph{enumi}}\refstepcounter
{enumi}\label{CorOveNov29.LowerBound(1)}%
\textup{(\ref{CorOveNov29.LowerBound(1)})} For all $\lambda\in\left[
\,\left(  \sqrt{5}-1\right)  /2,1\,\right)  $ we have%
\begin{equation}
\mu_{\lambda}\left(  \tau_{0}^{\left(  \lambda\right)  }\left(  X_{\lambda
}\right)  \cap\tau_{1}^{\left(  \lambda\right)  }\left(  X_{\lambda}\right)
\right)  \geq\frac{1}{3}. \label{eqOveNov29b.new1}%
\end{equation}

\refstepcounter{enumi}\label{CorOveNov29.LowerBound(2)}%
\textup{(\ref{CorOveNov29.LowerBound(2)})} For all $\lambda\in\left(
\,1/2,\left(  \sqrt{5}-1\right)  /2\,\right)  $, there is some $m$, depending
on $\lambda$, such that%
\begin{equation}
\lambda+\lambda^{2}+\lambda^{3}+\dots+\lambda^{m}\geq1,
\label{eqOveNov29b.new2}%
\end{equation}
and for such a choice of $m$ we have%
\begin{equation}
\mu_{\lambda}\left(  \tau_{0}^{\left(  \lambda\right)  }\left(  X_{\lambda
}\right)  \cap\tau_{1}^{\left(  \lambda\right)  }\left(  X_{\lambda}\right)
\right)  \geq\frac{1}{2^{m}-1}. \label{eqOveNov29b.new3}%
\end{equation}

\end{corollary}

\begin{remark}
\label{RemOve.GoldenBound}If $\lambda=\left(  \sqrt{5}-1\right)  /2$, the
number $m$ in \textup{(\ref{eqOveNov29b.new2})} may be taken to be $m=2$, and
in that case \textup{(\ref{eqOveNov29b.new2})} is \textquotedblleft%
$=$\textquotedblright.
\end{remark}

\begin{proof}
[Proof of Corollary \textup{\ref{CorOveNov29.LowerBound}}]\emph{Ad}
(\ref{CorOveNov29.LowerBound(1)}): An easy modification of the argument in the
proof of Proposition \ref{ProOveNov28.OneThird} shows that if $\lambda
^{2}+\lambda>1$, then (\ref{eqOveNov29b.new1}) holds.

\emph{Ad} (\ref{CorOveNov29.LowerBound(2)}): Let $\lambda\in\left(
1/2,\left(  \sqrt{5}-1\right)  /2\right)  $ be given. It follows from algebra
that if $m$ is sufficiently large, then (\ref{eqOveNov29b.new2}) must hold. We
pick $m$ to be the first number which gets the sum on the left-hand side in
(\ref{eqOveNov29b.new2}) $\geq1$.

Consider the following specific finite word $w$ in $\left\{  0,1\right\}
^{\text{finite}}$ given by%
\[
w=(\,0\underbrace{\,1\,1\,\dots\,1\,}_{m-1\text{ times}}\,)
\]
and generate more words as follows:%
\begin{align*}
&  (\,1\,\text{free}\,),\\
&  (\,w\,1\,\text{free}\,),\\
&  (\,w\,w\,1\,\text{free}\,),\\
&  (\,w\,w\,w\,1\,\text{free}\,),\\
&  \text{etc.,}%
\end{align*}
where \textquotedblleft free\textquotedblright\ means unrestricted strings of
bits. As before, the resulting sequence of subsets in $\Omega$ is disjoint.
Then%
\[
\mu_{\lambda}\left(  \tau_{1}^{\left(  \lambda\right)  }\left(  X_{\lambda
}\right)  \right)  \geq2^{-1}+2^{-m-1}+2^{-2m-1}+2^{-3m-1}+\cdots=2^{-1}%
\frac{1}{1-2^{-m}}=\frac{2^{m-1}}{2^{m}-1}.
\]
The argument from the proposition yields%
\[
\mu_{\lambda}\left(  \tau_{0}^{\left(  \lambda\right)  }\left(  X_{\lambda
}\right)  \cap\tau_{1}^{\left(  \lambda\right)  }\left(  X_{\lambda}\right)
\right)  \geq\frac{2^{m}}{2^{m}-1}-1=\frac{1}{2^{m}-1},
\]
which is the assertion (\ref{eqOveNov29b.new3}).
\end{proof}

\emph{Cascade approximation}. Each function $F_{n}^{\left(  \lambda\right)  }$
from the approximation to the cumulative distribution in Proposition
\ref{ProOveNov26.1} has a finite set of node points $N_{n}(\lambda)$, and
\[
\#\,N_{n}(\lambda)\leq2^{n}\text{\qquad for all }n;
\]
but for fixed $\lambda$, the configuration of points in $N_{n}(\lambda)$ can
be complicated for $n>2$ and large; and each set $N_{n}(\lambda)$ also depends
on the particular numerical choice of a value for $\lambda$.

This is borne out in the cascade of figures 
(see Figure \ref{FigOve.FlambdaCascade})
made for $\lambda=(\sqrt{5}-1)/2$. We have included pictures of $F_{1}%
^{\left(  \lambda\right)  }$, $F_{2}^{\left(  \lambda\right)  }$, $\dots$, up
to $F_{4}^{\left(  \lambda\right)  }$. As sketched in Remark
\ref{RemOveNov26.pound1}, the reason is that the detailed configuration and
the multiplicities in the sets $N_{n}(\lambda)$ of node points are reflected
in the progression of graphs of the functions $F_{n}^{\left(  \lambda\right)
}(\,\cdot\,)$. This fine structure only becomes visible for large $n>2$.

\begin{figure}[ptb]
\setlength{\unitlength}{0.47\textwidth}
\begin{picture}(2.118,1.5)(-0.25,-0.25)
\put(0,0){\includegraphics[bb=88 4 376 253, width=2\unitlength]{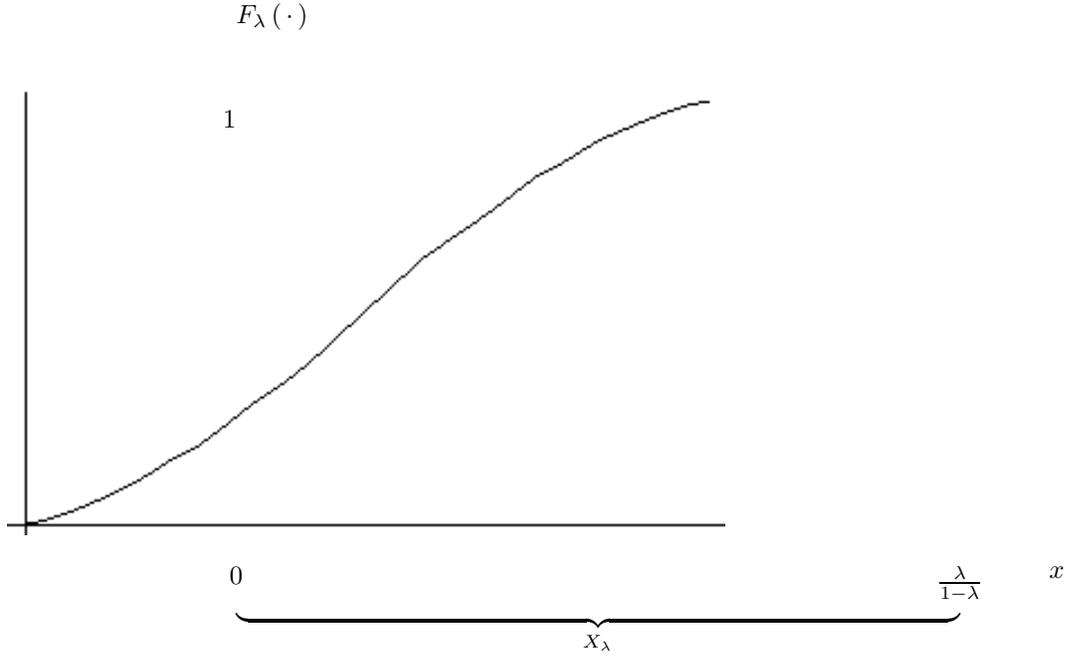}}
\put(0,1.2){\makebox(0,0)[bl]{$F_{\lambda}\left(\,\cdot\,\right)$}}
\put(0,1){\makebox(0,0)[r]{$1$}}
\put(0,0){\makebox(0,0)[t]{$0$}}
\put(1.618,0){\makebox(0,0)[t]{$\frac{\lambda}{1-\lambda}$}}
\put(1.818,0){\makebox(0,0)[tl]{$x$}}
\put(0.809,-0.1){\makebox(0,0)[t]{$\underbrace{\rule{1.618\unitlength}{0pt}}_{X_{\lambda}}$}}
\end{picture}
\caption{The cumulative distribution of $\pi_{\lambda}\left(  \,\cdot
\,\right)  $, where $\lambda = (\sqrt{5}-1)/2$. Caution: A closed formula for $F_{\lambda}\left(  \,\cdot
\,\right)  $ is not known. But using the second formula in (\ref{eqOve.1}) and
(\ref{eqOve.2}), the reader may check that for fixed $\lambda$, the cumulative
distribution $F$ ($= F_{\lambda}$) satisfies the scaling identity $F(x) =
\frac{1}{2}( F(x/\lambda) + F((x-\lambda)/\lambda))$.}%
\label{FigOve.Flambda}%
\end{figure}

In Figure \ref{FigOve.Flambda} above, we summarize the distribution of
$\pi_{\lambda}\left(  \,\cdot\,\right)  $, i.e., the function $F_{\lambda
}\left(  \,\cdot\,\right)  $ in (\ref{eqOve.2}). By \cite{Sol95}, $F_{\lambda
}\left(  \,\cdot\,\right)  $ is only known to have $L^{1}$-$\mathrm{a.e.}$
derivative, or $L^{1}$-density, for $\mathrm{a.e.}\;\lambda$ in $1/2<\lambda
<1$.

\begin{figure}
\setlength{\unitlength}{0.90bp}
\begin{picture}(252,117)(0,-45)
\put(0,0){\includegraphics[bb=88 4 340 76,width=252\unitlength]{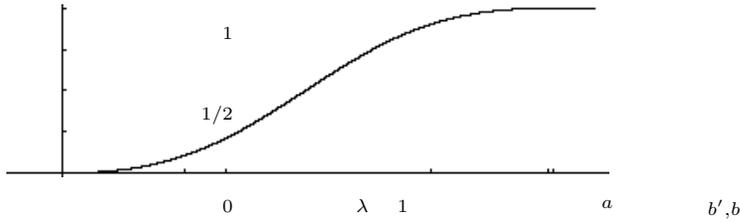}}
\put(21,70.4){\makebox(0,0)[r]{$\scriptstyle1$}}
\put(21,36){\makebox(0,0)[r]{$\scriptstyle1/2$}}
\put(21,0.5){\makebox(0,0)[tr]{$\scriptstyle0$}}
\put(75,0.5){\makebox(0,0)[t]{$\scriptstyle\lambda$}}
\put(92.1111,0.5){\makebox(0,0)[t]{$\scriptstyle1$}}
\put(177.667,0.5){\makebox(0,0)[t]{$\scriptstyle a$}}
\put(227.971,0.5){\makebox(0,0)[t]{$\scriptstyle b^{\prime},b^{\phantom{\prime}}$}}
\end{picture}
\caption{$\lambda = 3/4$}
\label{Figfplotabcdefg}
\end{figure}

\begin{figure}
\setlength{\unitlength}{0.90bp}
\begin{picture}(360,117)(0,-45)
\put(0,0){\includegraphics[bb=88 4 448 76,width=360\unitlength]{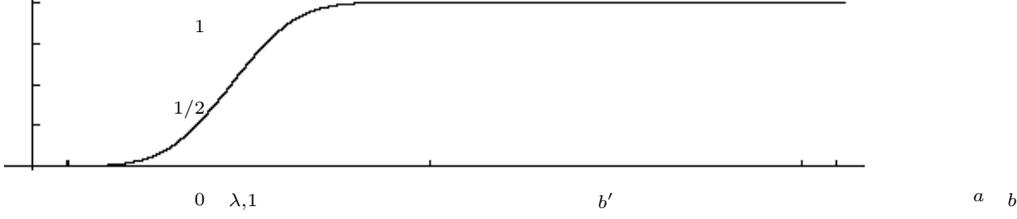}}
\put(10,70.4){\makebox(0,0)[r]{$\scriptstyle1$}}
\put(10,36){\makebox(0,0)[r]{$\scriptstyle1/2$}}
\put(10,0.5){\makebox(0,0)[tr]{$\scriptstyle0$}}
\put(26,0.5){\makebox(0,0)[t]{$\scriptstyle\lambda,1$}}
\put(334,0.5){\makebox(0,0)[t]{$\scriptstyle a$}}
\put(177.939,0.5){\makebox(0,0)[t]{$\scriptstyle b'$}}
\put(348,0.5){\makebox(0,0)[t]{$\scriptstyle b$}}
\end{picture}
\caption{$\lambda=23/24$}
\label{Figfplothijk}
\end{figure}

         The differences in cascade approximation when the value of $\lambda$
varies is illustrated by Figures \ref{FigOve.FlambdaCascade}, 
\ref{FigOve.Flambda}, \ref{Figfplotabcdefg}, and \ref{Figfplothijk}.  
We have already outlined that $\lambda = (\sqrt{5}-1)/2$ is special.  
Figure \ref{FigOve.FlambdaCascade}
illustrates the cascades from
Proposition \ref{ProOveNov26.1}, 
$F_0^{\left(  \lambda\right)  } \rightarrow F_1^{\left(  \lambda\right)  } \rightarrow \ldots \rightarrow F_7^{\left(  \lambda\right)  }$,
with each step representing a
$\lambda$ scaled subdivision.  Figure \ref{FigOve.Flambda} represents
the ``idealized'' limit of the iteration.  Since there are only very few rigorous results in the literature for specific values of $\lambda$ in the open interval
$(1/2, 1)$ (see Section \ref{Int}), we have included Mathematica sketches of of $F_{\lambda}$ for two chosen values of $\lambda$ in Figures \ref{Figfplotabcdefg} and \ref{Figfplothijk}.


Our next result gives the two Radon--Nikodym derivatives in the case of
(\ref{eqOve.1}) for $\lambda$ fixed, i.e., $1/2<\lambda<1$ is chosen. The
measure $\mu=\mu_{\lambda}$ is chosen such that $\mu_{\lambda}\left(
X_{\lambda}\right)  =1$. Then set%
\begin{equation}
\varphi_{i}:=\frac{d\mu\circ\tau_{i}^{-1}}{d\mu},\qquad i=0,1.
\label{eqOve.2bis}%
\end{equation}


\begin{proposition}
\label{ProOve.RadNik}The two Radon--Nikodym derivatives $\varphi_{0}$ and
$\varphi_{1}$ from \textup{(\ref{eqOve.2bis})} are given by the formulas%
\begin{align}
\frac{1}{2}\,\varphi_{0}\left(  x\right)   &  =\left\{
\begin{aligned} &1&&\text{ if }0\leq x<\lambda,\vphantom{\frac{\lambda^{2}}{1-\lambda}}\\ &F\left( \frac{\lambda^{2}-x\left( 1-\lambda\right) }{\left( 1-\lambda\right) \left( 2\lambda-1\right) }\right) &&\text{ if }\lambda\leq x\leq\frac{\lambda^{2}}{1-\lambda},\\ &0&&\text{ if }\frac{\lambda^{2}}{1-\lambda }<x\leq\frac{\lambda}{1-\lambda}, \end{aligned}\right.
\label{eqOve.3}\\%
\intertext{and}%
\frac{1}{2}\,\varphi_{1}\left(  x\right)   &  =\left\{
\begin{aligned} &0&&\text{ if }0\leq x<\lambda,\vphantom{\frac{\lambda^{2}}{1-\lambda}}\\ &\rlap{$\displaystyle F\left(  \frac{\left(  x-\lambda\right)  }{\left(  2\lambda
-1\right)  }\right)  $}\phantom{F\left( \frac{\lambda^{2}-x\left( 1-\lambda\right) }{\left( 1-\lambda\right) \left( 2\lambda-1\right) }\right) }&&\text{ if }\lambda\leq x\leq\frac{\lambda^{2}}{1-\lambda },\\ &1&&\text{ if }\frac{\lambda^{2}}{1-\lambda}<x\leq\frac{\lambda}{1-\lambda }. \end{aligned}\right.
\label{eqOve.4}%
\end{align}

\end{proposition}

\begin{proof}
The result follows from Corollary \ref{CorCon.EquiMe} and from the present
discussion of Example \ref{ExaCon.xplus1}.
\end{proof}


\begin{figure}[ptb]
\setlength{\unitlength}{0.23\textwidth} \begin{minipage}{0.5\textwidth}
\begin{picture}(2.118,2.5)(-0.25,-0.3)
\put(0,0){\includegraphics[height=2\unitlength, width=1.65\unitlength]{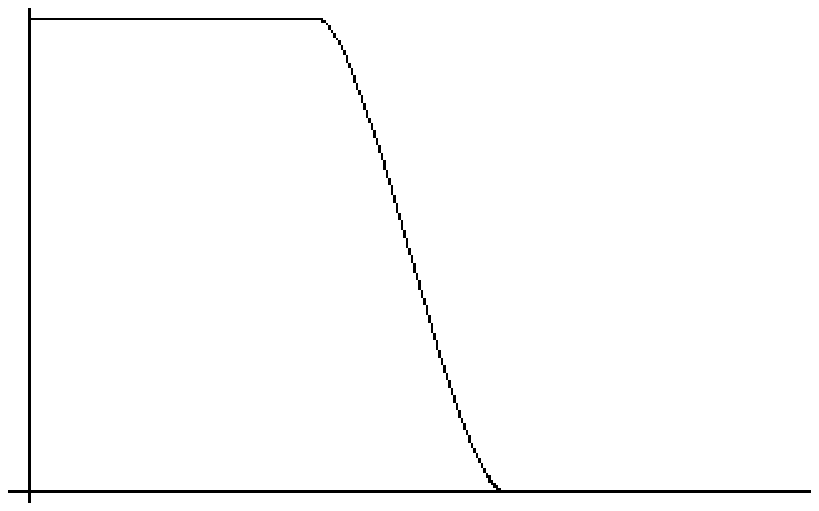}}
\put(-0.1,1.2){\makebox(0,0)[r]{$\varphi_{0}$}}
\put(0,2){\makebox(0,0)[r]{$2$}}
\put(0.05,0){\makebox(0,0)[tr]{$X_{\lambda}:0$}}
\put(1.618,0){\makebox(0,0)[t]{$\frac{\lambda}{1-\lambda}$}}
\put(1,0){\makebox(0,0)[t]{$\frac{\lambda^{2}}{1-\lambda}$}}
\put(0.618,0){\makebox(0,0)[t]{$\lambda$}}
\put(0.809,-0.2){\makebox(0,0)[t]{\textsc{overlap}}}
\end{picture}
\caption{The Radon--Nikodym derivative $\varphi_{0}$.}%
\label{FigOve.phi0}
\end{minipage}\begin{minipage}{0.5\textwidth}
\begin{picture}(2.118,2.5)(-0.25,-0.3)
\put(0,0){\includegraphics[height=2\unitlength, width=1.65\unitlength]{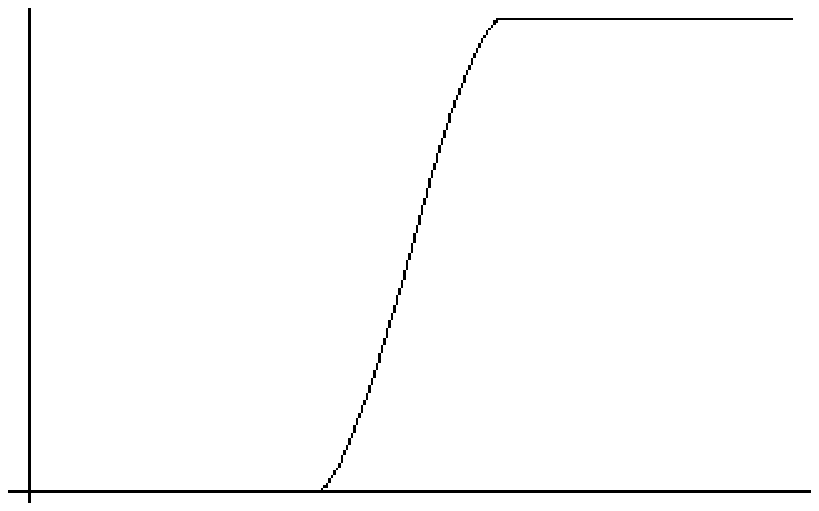}}
\put(-0.1,1.2){\makebox(0,0)[r]{$\varphi_{1}$}}
\put(0,2){\makebox(0,0)[r]{$2$}}
\put(0.05,0){\makebox(0,0)[tr]{$X_{\lambda}:0$}}
\put(1.618,0){\makebox(0,0)[t]{$\frac{\lambda}{1-\lambda}$}}
\put(1,0){\makebox(0,0)[t]{$\frac{\lambda^{2}}{1-\lambda}$}}
\put(0.618,0){\makebox(0,0)[t]{$\lambda$}}
\put(0.809,-0.2){\makebox(0,0)[t]{\textsc{overlap}}}
\end{picture}
\caption{The Radon--Nikodym derivative $\varphi_{1}$.}%
\label{FigOve.phi1}
\end{minipage}\end{figure}

\begin{remark}
\label{RemOve.Sketch}For the convenience of the reader, we have sketched the
two functions $\varphi_{0}$ and $\varphi_{1}$ in Figures
\textup{\ref{FigOve.phi0}} and \textup{\ref{FigOve.phi1}} above.\medskip

\noindent\textit{Details for \textup{(\ref{eqOve.4})}, Figure
\textup{\ref{FigOve.phi1}}.} Using \textup{(\ref{eqCon.21})},
\textup{(\ref{eqOve.2})}, Corollary \textup{\ref{CorCon.EquiMe}}, and
conditional probabilities, we get for all $S\in\mathcal{B}$ \textup{(}Borel
subsets of $X_{\lambda}$\textup{)} the formula%
\begin{equation}
\mu\left(  \tau_{1}^{-1}\left(  S\right)  \right)  =\int_{S}P\left(  \left\{
\pi_{\lambda}\circ\sigma_{1}\leq x\right\}  \right)  \,d\mu\left(  x\right)  ,
\label{eqRemOve.Sketchpound}%
\end{equation}
where the function under the integral sign in
\textup{(\ref{eqRemOve.Sketchpound})} coincides with the expression in
\textup{(\ref{eqOve.4})}.\qed
\end{remark}

\begin{remark}
\label{RemOve.Chomu}As above, let $1/2<\lambda<1$, and $X_{\lambda}=\left[
\,0,\lambda/\left(  1-\lambda\right)  \,\right]  $. Then $\tau_{0}\left(
X_{\lambda}\right)  =\left[  \,0,\lambda^{2}/\left(  1-\lambda\right)
\,\right]  $ and $\tau_{1}\left(  X_{\lambda}\right)  =\left[  \,\lambda
,\lambda/\left(  1-\lambda\right)  \,\right]  $; see Figures
\textup{\ref{FigOve.phi0}} and \textup{\ref{FigOve.phi1}}. We include this
note to stress that the Radon--Nikodym derivatives are sensitive to the choice
of $\mu$. Take for example $\mu={}$Lebesgue measure on $X_{\lambda}$. Then an
easy computation yields%
\[
\frac{d\mu\circ\tau_{i}^{-1}}{d\mu}=\frac{1}{\lambda}\chi_{\tau_{i}\left(
X_{\lambda}\right)  },\qquad i=0,1;
\]
but the corresponding $\mathbb{F}$ is then \emph{not} a column isometry.
\end{remark}

\begin{proposition}
\label{ProOve.RanPro}Continuing the example in Proposition
\textup{\ref{ProOve.RadNik}}. If we choose $\mu$ to be the equilibrium
measure, then the isometry $\mathbb{F}=%
\begin{pmatrix}
F_{0}\\
F_{1}%
\end{pmatrix}
\colon L^{2}\left(  \mu\right)  \rightarrow L^{2}\left(  \mu\right)  _{2}$
yields the following formula for the range-projection in $L^{2}\left(
\mu\right)  _{2}$:%
\begin{equation}
\mathbb{FF}^{\ast}=%
\begin{pmatrix}
F_{0}F_{0}^{\ast} & F_{0}F_{1}^{\ast}\\
F_{1}F_{0}^{\ast} & F_{1}F_{1}^{\ast}%
\end{pmatrix}
=\frac{1}{2}%
\begin{pmatrix}
\varphi_{0}\circ\tau_{0} & \left(  \varphi_{1}\circ\tau_{0}\right)  T_{-1}\\
\left(  \varphi_{0}\circ\tau_{1}\right)  T_{1} & \varphi_{1}\circ\tau_{1}%
\end{pmatrix}
, \label{eqOve.5}%
\end{equation}
where the composite functions $\varphi_{i}\circ\tau_{j}$ serve as
multiplication operators in $L^{2}\left(  \mu\right)  $, while the two other
operators making up the entries in \textup{(\ref{eqOve.5})} are%
\[
\left(  T_{\pm1}f\right)  \left(  x\right)  =f\left(  x\pm1\right)  ,\qquad
f\in L^{2}\left(  \mu\right)  .
\]
As before $\lambda$ is fixed such that $1/2<\lambda<1$, and $\mu=\mu_{\lambda
}$ is the equilibrium measure.
\end{proposition}

\begin{proof}
The result follows from a computation, and an application of Remark
\ref{RemMul.ComOp} and Lemma \ref{LemMul.AdjForm}.
\end{proof}


\begin{corollary}
\label{CorOve.OrtCom}Continuing the example from Propositions
\textup{\ref{ProOve.RadNik}} and \textup{\ref{ProOve.RanPro}}, we note that
the function
\begin{equation}
\mathbf{f}=%
\begin{pmatrix}
f_{0}\left(  x\right) \\
f_{1}\left(  x\right)
\end{pmatrix}
=%
\begin{pmatrix}
\varphi_{1}\circ\tau_{0}\\
-\varphi_{0}\circ\tau_{1}%
\end{pmatrix}
\in L^{2}\left(  \mu\right)  _{2} \label{eqOve.6}%
\end{equation}
is nonzero, and it is in the orthogonal complement of the range of the column
isometry $\mathbb{F}$.
\end{corollary}

\begin{proof}
Using (\ref{eqOve.5}), a computation shows that $\mathbf{f}$ in (\ref{eqOve.6}%
) satisfies $\mathbb{FF}^{\ast}\mathbf{f}=0$; and moreover that $\mathbf{f}%
=\left(
\begin{smallmatrix}
f_{0}\\
f_{1}%
\end{smallmatrix}
\right)  $ is nonzero in $L^{2}\left(  \mu\right)  _{2}$.
\end{proof}

\section{\label{Sie}Essential overlap and gaps for Sierpinski constructions in
$\mathbb{R}^{2}$}

This section includes a 2D variant of the 1D examples from Section \ref{Ove}
above. The case of 2D is interesting and different in that the fractal
features become apparent both for the associated Hutchinson measures, as well
as for their support. Contrast: In the best known gap fractal, the Cantor set,
the middle thirds are omitted, and as a result the familiar cascading
gap-subintervals emerge. As we saw in Example \ref{ExaCon.xplus1}, in 1D
(\textsc{Case 3}) when $\lambda$ is adjusted, $\lambda> 1/2$ so as to create
$\mu_{\lambda}$-essential overlap for the infinite convolution IFS, then the
omitted middles disappear, and the resulting $X_{\lambda}$ is simply an
interval. Not so for the analogous 2D construction! As we see, in 2D overlap
and gaps may coexist!

Recall that the attractor for a contractive IFS in a complete metric space is
equal to the support of the corresponding Hutchinson measure.

In the next result, we show that when the same procedure from Proposition
\ref{ProOveNov26.1} (and (\ref{eqOve.1})--(\ref{eqOve.2})) is extended to 2D
we get a one-parameter family of fractals with gaps, and essential overlap at
the same time. So in the context of (\ref{eqCon.2}) and (\ref{eqCon.3}), the
ambient space $Y=\mathbb{R}^{2}$, the weights in (\ref{eqCon.3}) are
$p_{i}=1/3$, the number $\lambda$ will be in the open interval $\left(
1/2,2/3\right)  $, and the attractor $X=X_{\lambda}$ will be a Sierpinski
fractal with essential overlap and gaps. Its Hausdorff dimension $H_{\lambda}$
will be
\[
H_{\lambda}=-\frac{\log3}{\log\lambda}.
\]
This Sierpinski fractal${}=X_{\lambda}$ is sketched below in Figures
\ref{FigSie.4}, \ref{FigSie.1}, \ref{FigSie.3}, and \ref{FigSie.2} for the cases $\lambda = 11/20$, $\lambda=\left(  \sqrt
{5}-1\right)  /2$, $\lambda = 13/20$, and $\lambda={3/4}$, respectively.


\begin{figure}
\setlength{\unitlength}{1bp}
\bigskip
\bigskip
\begin{picture}(324,162)(0,-45)
\put(0,0){\includegraphics[width=153\unitlength]{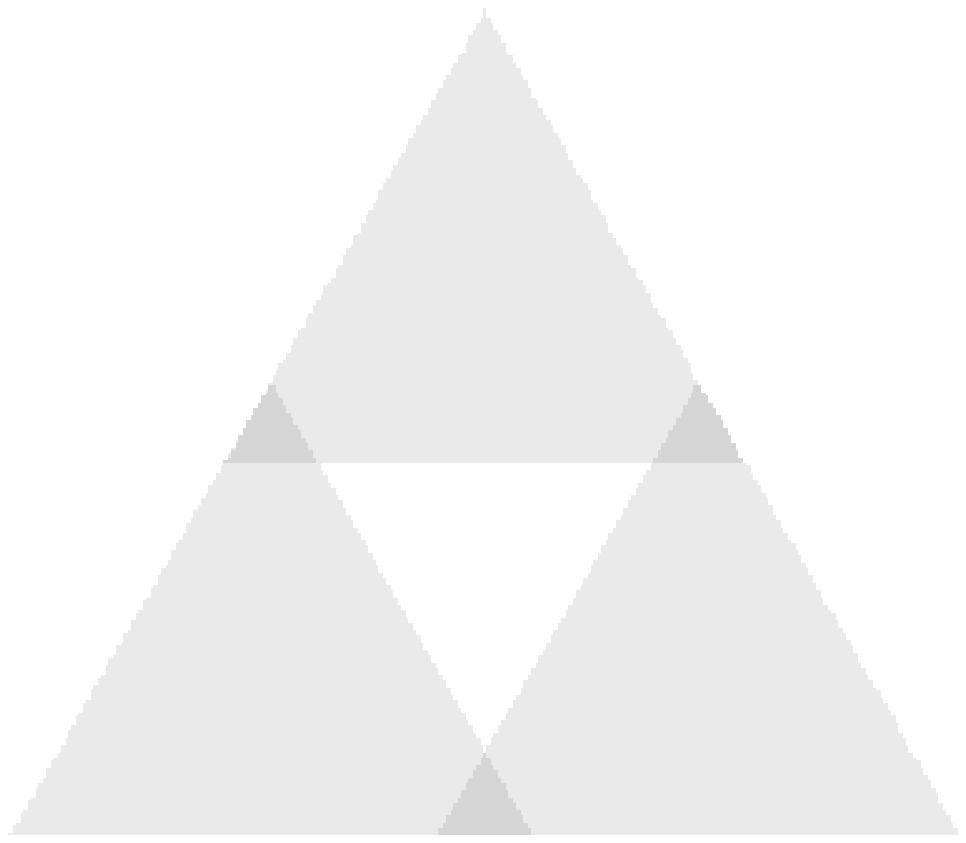}}
\put(76,-13){\makebox(0,0)[t]{(a). $n=1$}}

\put(171,0){\includegraphics[width=153\unitlength]{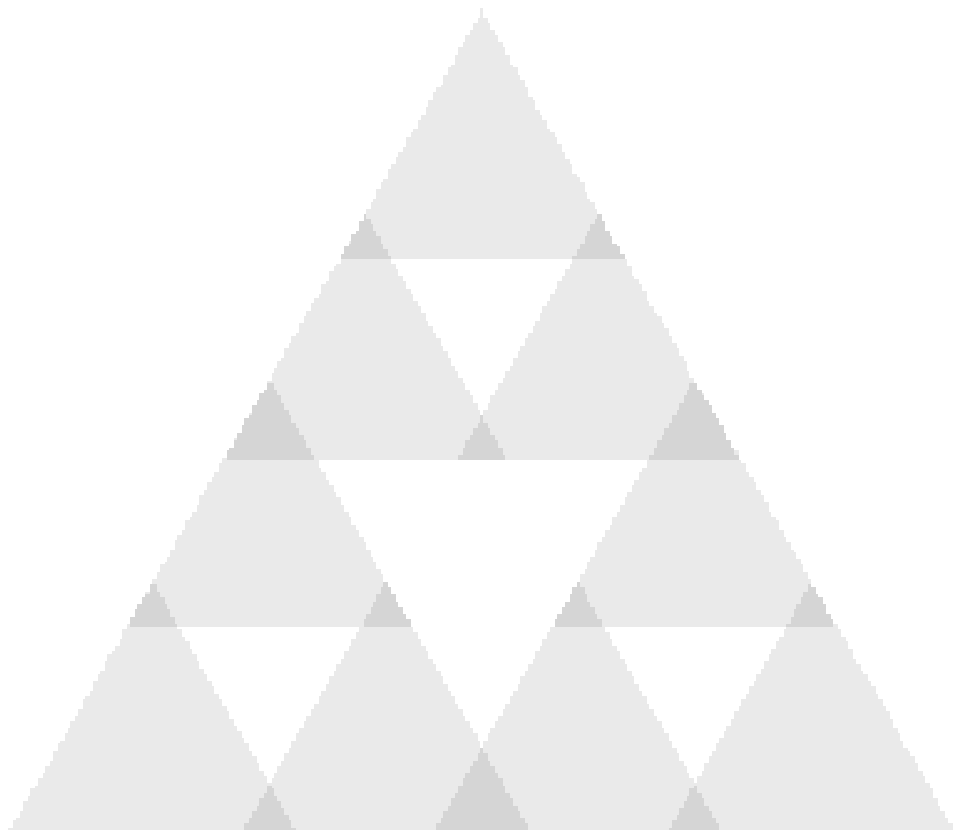}}
\put(247,-13){\makebox(0,0)[t]{(b). $n=2$}}
\end{picture}

\caption{The first two iterations of $X_{\lambda}$ for $\lambda = 11/20$.
}
\label{FigSie.4}
\end{figure}

Let $\lambda$ be fixed in the open interval $\left(  1/2, 2/3\right)  $. The
case $\lambda= 1/2$ is the familiar Sierpinski gasket in 2D.

Recursion: By analogy to the middle-third Cantor-set construction, start with
a triangle $T_{\lambda}$ with vertices $\left(  0,0\right)  $, $\left(
\lambda/\left(  1 - \lambda\right)  \right)  u_{1}$, and $\left(
\lambda/\left(  1 - \lambda\right)  \right)  u_{2}$.

The three $\lambda$-scaled triangles $\tau_{i}\left(  T_{\lambda}\right)  $
are shaded in light grayscale, and they have pairwise overlaps as indicated in
the first iteration in Fig.\ \ref{FigSie.1}. These three pairwise overlaps are
the first smaller dark-shaded triangles. The recursion now continues.

Omitted triangles: The first gap, i.e., the first white interior triangle, is
the set-theoretic difference $G_{1}=T_{\lambda}\setminus\bigcup_{i}\tau
_{i}\left(  T_{\lambda}\right)  $. The fractal $X_{\lambda}$ now arises by
iteration just as in the familiar case of recursion for the middle-third
Cantor set. In the first step of the recursion, there is just one interior and
centered gap-triangle; it is inverted from the position of the initial
$T_{\lambda}$.

Also note that the size of the overlaps in the iteration is monotone
in $\lambda$, small when $\lambda$ is close to $1/2$, the usual
Sierpinski gasket.  In contrast, the omitted middles decrease
gradually with $\lambda$ and they collapse to points when $\lambda =
2/3$.  The nature of the overlaps changes at the value $\lambda =
(\sqrt{5}-1)/2$. We will discuss the nature of the overlaps in Section \ref{natureoverlap}.

\subsection{The 2D recursion}\label{recursion}

Initialize the recursion sequence at level $n=0$ with an inflated triangle
$T_{\lambda}$, inflation factor${}=\lambda/\left(  \lambda-1\right)  $, and
then generate the fractal $X_{\lambda}$ sequentially, $n=1,2,\dots$, by the
usual iteration limit, 
\[X_{\lambda}=\overline{\bigcap_{n=1}^{\infty}%
\bigcup_{w\text{: word of length }n}\tau_{w}\left(  T_{\lambda}\right)  }.\]
Here $\tau_{w}$ denotes an $n$-fold composition of the individual $\tau_{i}$
maps with the indices $I$ making up the word $w$, i.e., $w$ giving the address
of the respective \textquotedblleft small\textquotedblright\ triangles
$\tau_{w}\left(  T_{\lambda}\right)  $ for the $n$'th level iteration:
\begin{gather*}
\tau_{i}\left(  x\right)  =\lambda\left(  x+u_{i}\right)  ,\qquad
i=0,1,2,\;x\in\mathbb{R}^{2},\\
\tau_{w}=\tau_{i_{1}}\cdots\tau_{i_{n}}\text{\qquad if }w=\left(  i_{1}%
,\dots,i_{n}\right)  .
\end{gather*}


\begin{figure}
\setlength{\unitlength}{1bp}
\bigskip
\bigskip
\begin{picture}(324,162)(0,-45)
\put(0,0){\includegraphics[width=153\unitlength]{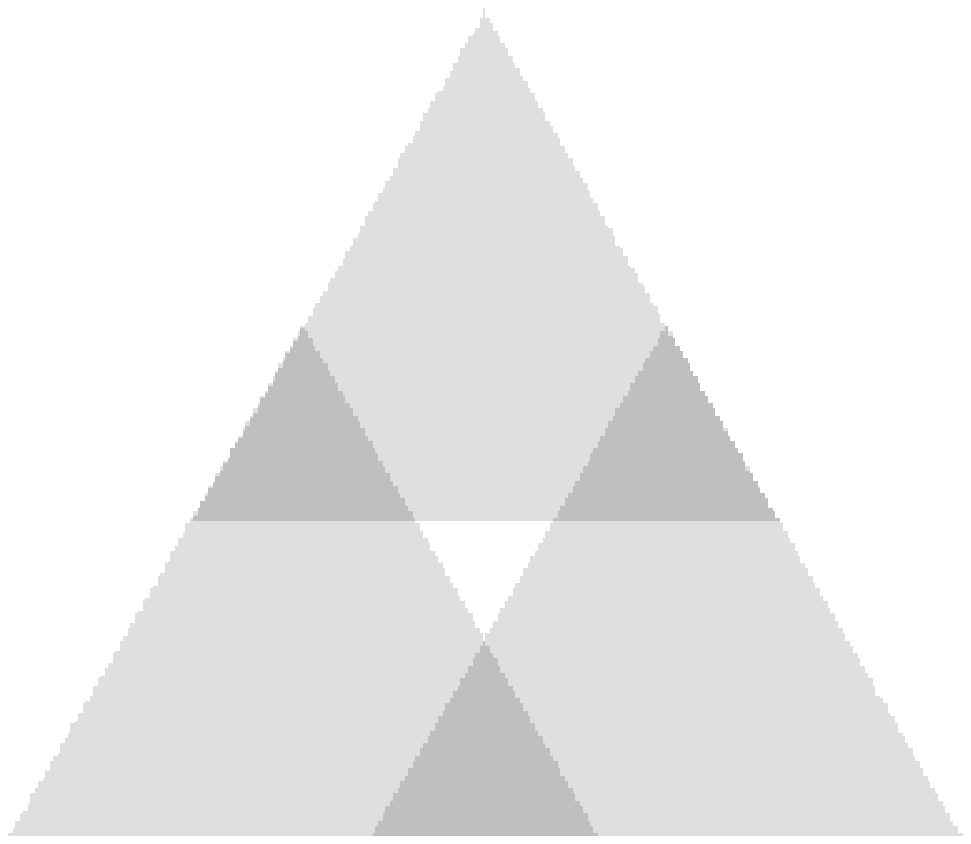}}
\put(76,-13){\makebox(0,0)[t]{(a). $n=1$}}

\put(171,0){\includegraphics[width=153\unitlength]{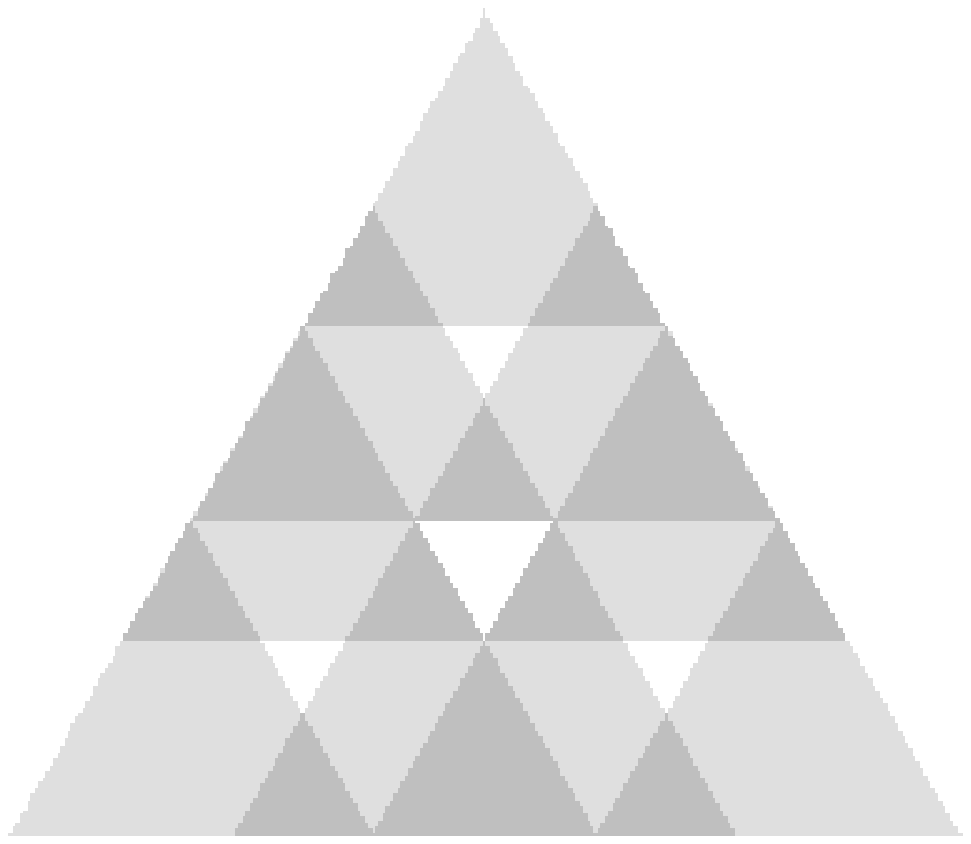}}
\put(247,-13){\makebox(0,0)[t]{(b). $n=2$}}
\end{picture}

\caption{The first two iterations of $X_{\lambda}$ for $\lambda = (\sqrt{5}-1)/2$.  See Example \ref{ExaSie.1}.}
\label{FigSie.1}
\end{figure}

\begin{example}
\label{ExaSie.1}\textbf{2D Sierpinski with overlap and gaps,} $1/2<\lambda
<2/3$\textbf{.} Let the vectors $u_{0}$, $u_{1}$, $u_{2}$ in $\mathbb{R}^{2}$
be given as follows:%
\begin{equation}
\left\{
\begin{aligned} u_{0}&=\left( 0,0\right) \text{,}\\u_{1}&=\left( 1,0\right) \text{, and}\\u_{2}&=\left( \frac{1}{2},\frac{\sqrt{3}}{2}\right) \text{, and set} \end{aligned}\right.
\label{eqSie.1}%
\end{equation}%
\begin{equation}
\Omega=\left\{  u_{0},u_{1},u_{2}\right\}  ^{\mathbb{N}} \label{eqSie.2}%
\end{equation}
with Bernoulli measure $P_{1/3}$ corresponding to the infinite-product measure
$p_{i}=1/3$, $i=0,1,2$, as in \textup{(\ref{eqCon.3})} and Corollary
\textup{\ref{CorCon.BernMe}}. We will denote by $\mu_{\lambda}$ the measure in
Corollary \textup{\ref{CorCon.EquiMe}}, and we set $X_{\lambda}%
:=\operatorname*{supp}\left(  \mu_{\lambda}\right)  $.

The IFS is%
\begin{equation}
\left\{
\begin{aligned}\tau_{0}^{\left( \lambda\right) }\left( x\right) &=\lambda x\text{,}\\\tau_{1}^{\left( \lambda\right) }\left( x\right) &=\lambda\left( x+u_{1}\right) \text{, and}\\\tau_{2}^{\left( \lambda\right) }\left( x\right) &=\lambda\left( x+u_{2}\right) \end{aligned}\right.
\label{eqSie.3}%
\end{equation}
for $x\in\mathbb{R}^{2}$. Here we make the restriction $1/2<\lambda<2/3$. As
noted in Section \textup{\ref{Con}},
\begin{equation}
X_{\lambda}=\bigcup_{i=0}^{2}\tau_{i}^{\left(  \lambda\right)  }\left(
X_{\lambda}\right)  \label{eqSie.4}%
\end{equation}
and $X_{\lambda}$ is the unique compact \textup{(}$\neq\varnothing$\textup{)}
solution to \textup{(\ref{eqSie.4})}.
\end{example}

Let $\Omega:=\{0, u_1, u_2\}^{\mathbb{N}}$, and let $P_{1/3}$ be the usual Bernoulli measure on $\Omega$ with equal and independent probabilities $(1/3, 1/3, 1/3)$.  The formula (\ref{eqCon.20}) for the random variable $\pi_{\lambda}: \Omega \rightarrow X_{\lambda}$ extends from 1D to 2D with the only modification that the coefficients $\omega_i$ from (\ref{eqCon.20}) now take values in the finite alphabet of vectors $\{ 0, u_1, u_2\}$.

Let $A_{i}^{\lambda}$, $i = 0, 1, 2$ be the three vertices in $X_{\lambda}$ (Figure \ref{FigSie.1}); i.e. $A_{0}^{\lambda} = (0,0)$, $A_{1}^{\lambda} = \frac{\lambda}{1-\lambda}u_1$, and $A_{2}^{\lambda} = \frac{\lambda}{1-\lambda} u_2$.  Note that for each $\lambda\in (1/2, 1)$, $X_{\lambda}$ is contained in the triangle $T_{\lambda}$ with the vertices $A_{i}^{\lambda}$, $i = 0, 1, 2$.  

Our first result concerns symmetry, and it is an immediate extension of Lemma \ref{LemOve.Symmetry} from the 1D case to the 2D case.  For each $i\in\{0, 1, 2\}$, let $S_{i}^{\lambda}(x)$ denote the equilateral triangle of side-length $x$ with vertex $A_{i}^{\lambda}$, which shares two sides with segments of sides in $T_{\lambda}$.  Then the argument from Lemma \ref{LemOve.Symmetry} shows that for each $x\in\mathbb{R}_{+}$, the three numbers
\begin{equation*}
P_{1/3}\{ \omega\in\Omega | \pi_{\lambda}(\omega) \in S_{i}^{\lambda}(x)\}
\end{equation*}
coincide.  Since $\mu_{\lambda} = P_{1/3}\circ \pi_{\lambda}^{-1}$ by Corollary \ref{CorCon.EquiMe}, we conclude in particular that the three numbers $\mu_{\lambda}(\tau_{i}^{(\lambda)}(X_{\lambda})$ agree for $i = 0, 1, 2$.

For $i\neq j$, set 
\[OV_{ij}^{\lambda}:=\tau_{i}^{(\lambda)}(X_{\lambda})\cap \tau_{j}^{(\lambda)}(X_{\lambda}).\]  It follows that 
\[ \mu_{\lambda}(OV_{01}^{\lambda}) = \mu_{\lambda}(OV_{02}^{\lambda}) = \mu_{\lambda}(OV_{12}^{\lambda}) .\]

\begin{proposition}
\label{ProSie.1}
\setcounter{enumi}{0}\renewcommand{\theenumi}{\alph{enumi}}\refstepcounter
{enumi}\label{ProSie.1(1)}\textup{(\ref{ProSie.1(1)})} For $\lambda\in\left(
1/2,2/3\right)  $, the fractal $X_{\lambda}$ has Hausdorff dimension%
\begin{equation}
H_{\lambda}=-\frac{\log3}{\log\lambda}. \label{eqSie.5}%
\end{equation}
It has essential overlap%
\begin{equation}
\mu_{\lambda}\left(  \tau_{i}^{\left(  \lambda\right)  }\left(  X_{\lambda
}\right)  \cap\tau_{j}^{\left(  \lambda\right)  }\left(  X_{\lambda}\right)
\right)  >0\text{\qquad for }i\neq j \label{eqSie.6}%
\end{equation}
and the $\mu_{\lambda}$-measure of the pairwise overlaps is independent of the
pair $\left(  i,j\right)  $ with $i\neq j$.

\refstepcounter{enumi}\label{ProSie.1(2)}\textup{(\ref{ProSie.1(2)})} Setting%
\begin{equation}
\pi_{\lambda}\left(  \omega\right)  =\sum_{i=1}^{\infty}\omega_{i}\lambda^{i},
\label{eqSie.7}%
\end{equation}
$\omega=\left(  \omega_{i}\right)  _{1}^{\infty}\in\Omega$, we get a
vector-valued random variable, and%
\[
\mu_{\lambda}=P_{1/3}\circ\pi_{\lambda}^{-1}%
\]
holds, where $\pi_{\lambda}\colon\Omega\rightarrow X_{\lambda}$ is the
encoding mapping.

\refstepcounter{enumi}\label{ProSie.1(3)}\textup{(\ref{ProSie.1(3)})}%
\[
\tau_{0}^{\left(  \lambda\right)  }\left(  X_{\lambda}\right)  \cap\tau
_{1}^{\left(  \lambda\right)  }\left(  X_{\lambda}\right)  \cap\tau
_{2}^{\left(  \lambda\right)  }\left(  X_{\lambda}\right)  =\varnothing.
\]

\end{proposition}


\begin{figure}
\setlength{\unitlength}{1bp}
\bigskip
\bigskip
\begin{picture}(324,162)(0,-45)
\put(0,0){\includegraphics[width=153\unitlength]{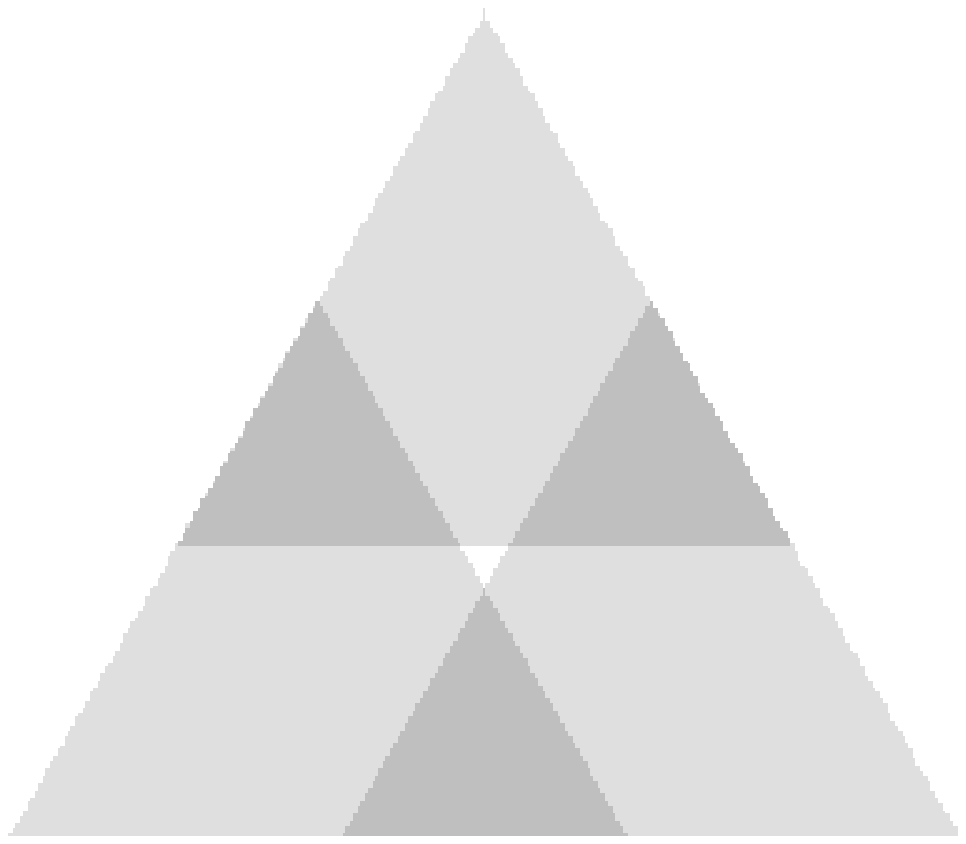}}
\put(76,-13){\makebox(0,0)[t]{(a). $n=1$}}

\put(171,0){\includegraphics[width=153\unitlength]{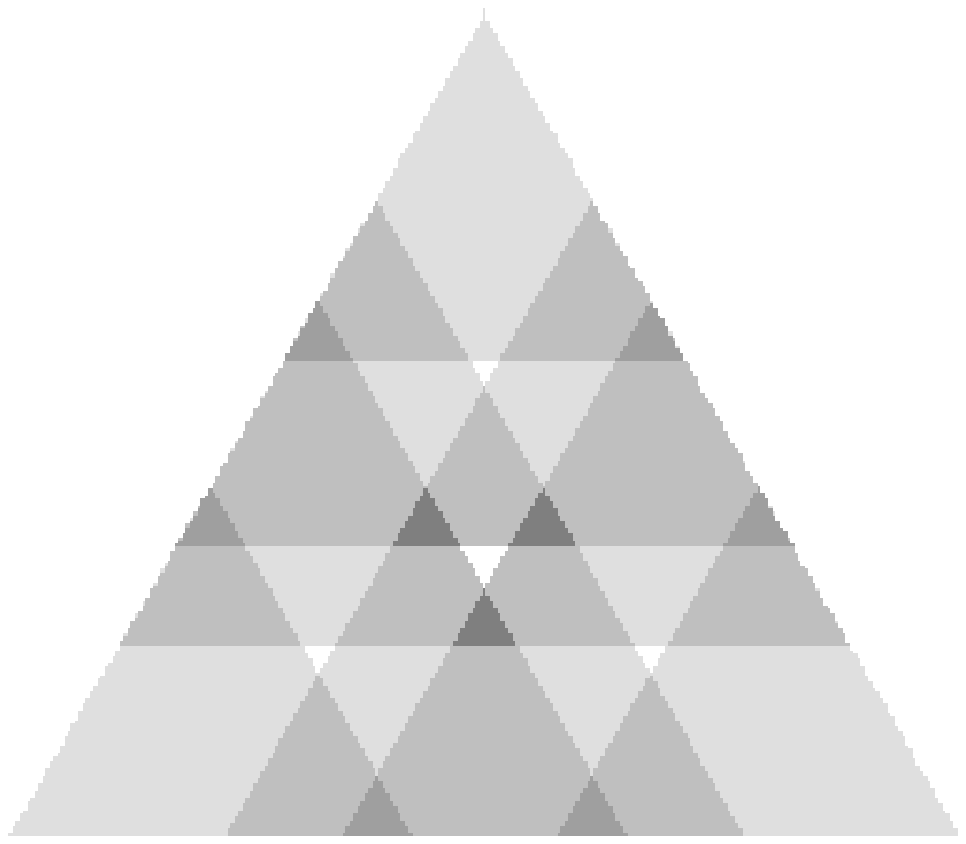}}
\put(247,-13){\makebox(0,0)[t]{(b). $n=2$}}
\end{picture}

\caption{The first two iterations of $X_{\lambda}$ for $\lambda = \frac{13}{20}$.  Note that $\frac{2}{3} - \frac{13}{20} = \frac{1}{60}$, and the gaps are very small.}
\label{FigSie.3}
\end{figure}

\begin{proof}
Since the proof is essentially contained in the previous sections, we shall be brief.

\emph{Ad} (\ref{ProSie.1(1)}): The formula (\ref{eqSie.5}) in
(\ref{ProSie.1(1)}) for the Hausdorff dimension follows from the arguments in
\cite{ChIv06}.   

We begin with the case $\lambda = (\sqrt{5}-1)/2$, and for the moment, we will drop $\lambda$ from the notation.  Our claim is that \[ \mu(OV_{01}) = \mu(OV_{02}) = \mu(OV_{12}) = \frac{1}{24} .\]
To see this, we introduce the Bernoulli space $\Omega$ and consider the cylinder sets in $\Omega$ indexed by finite words in the alphabet $\{u_0, u_1, u_2\}$, $u_0 = 0$.  If $w = (u_{i_1}\,u_{i_2} \ldots u_{i_n})$, set \[\Omega(w) := \{ \omega \in \Omega | \omega_1 = u_{i_1}, \ldots, \omega_n = u_{i_n}\},\] and note that $P_{1/3}(\Omega(w)) = 3^{-n}$. 

Set $w = (u_0 u_1) = (0\,u_1)$, and consider the following sequence of disjoint cylinder sets: \[ \Omega(u_1), \Omega(w\,u_1), \Omega(w\,w\, u_1), \ldots, \Omega(\underbrace{w\,w\cdots w}_{\text{ $k$ times}}\,u_1), \ldots \,.\]  The argument from Section \ref{Ove} shows that
\begin{equation*}
\mu(\tau_1(X))  = \sum_{k = 0}^{\infty} P_{1/3}\Bigl(\Omega(\underbrace{w\,w\cdots w}_{\text{ $k$ times}}\,u_1)\Bigr) = \sum_{k=0}^{\infty} \Bigl(\frac{1}{3}\Bigr)^{2k+1} = \frac{3}{8}.
\end{equation*}

Using the symmetry argument from above, we also have
\begin{equation*}
1 = \mu\Bigl( \cup_{i} \tau_i(X) \Bigr) = 3\mu(\tau_1(X)) - 3 \mu(OV_{01}) = \frac{9}{8} - 3\mu(OV_{01}).
\end{equation*}
The desired result $\mu(OV_{01}) = 1/24$ follows.  

Starting at $\lambda_1 = (\sqrt{5}-1)/2$ we see that the function $\lambda\mapsto \mu_{\lambda}(OV_{01}^{\lambda})$ is increasing.  Hence, to prove (\ref{eqSie.6}), we need only establish a lower bound for values of $\lambda$ in the open interval $(1/2, (\sqrt{5}-1)/2)$.  But this can be done \textit{mutatis mutandis} as in the proof of Corollary \ref{CorOveNov29.LowerBound}; see (\ref{eqOveNov29b.new3}).  If $\lambda\in(1/2, (\sqrt{5}-1)/2)$, determine $m\in\mathbb{N}$ as in (\ref{eqOveNov28.poundprimeprimeprime}).  Then it follows that
\[3 \mu_{\lambda}(OV_{01}^{\lambda}) \geq \frac{1}{3^m - 1}.\]
\emph{Ad} (\ref{ProSie.1(2)}): The proof of (\ref{eqSie.6}) in
(\ref{ProSie.1(2)}) is based on symmetry considerations (Lemma
\ref{LemOve.Symmetry}) extended from 1D to 2D, as well as the estimates in
Proposition \ref{ProOveNov28.OneThird}, Remarks \ref{RemOve.MeasureForMeasure}%
, and Corollary \ref{CorOveNov29.LowerBound}.

\emph{Ad} (\ref{ProSie.1(3)}): To see that the triple overlap is empty,
calculate the distances between pairwise overlaps to the third sub-partition.
\end{proof}

\nocite{ChIv06}

We conclude this section with the following open question: for what
values of $\lambda$ in the interval $(2/3, 1)$ is $\mu_{\lambda}$
absolutely continuous with respect to the $2$-dimensional Lebesgue
measure?  There are two pieces of partial evidence for absolute
continuity of the two-dimensional $\mu_{\lambda}$ for a.e.  $\lambda
\in (2/3 , 1)$:

\begin{enumerate}[(1)]
\item Since the interior gaps close at $\lambda = 2/3$, so that for
$\lambda \geq 2/3$, $X_\lambda = T_{\lambda}$ (the closed triangle),
following the 1D analogy, it seems reasonable to expect the a.e.
conclusion in this range of $\lambda$.

\item One of the proofs \cite{PeSo96} in the literature for the 1D case
introduces a clever Fubini-Tonelli argument with a function in several
variables with $\lambda$ as one of the integration variables. Using a density
argument, one then gets finiteness of a corresponding key functions for a.e.
$\lambda$ in the interval, and this in turn (following the 1D analogy) is
likely to yield an expression in 2D for the Radon-Nikodym derivative for
those values of $\lambda$.  

In particular, for $x\in T_{\lambda}$, let $L_{\lambda}^{\ell}(x)$ be the equilateral triangle centered at $x$ with side length $\ell$ and area $\frac{\sqrt{3}}{4}\ell^2$.   Define 
\[ \underline{D}(\mu_{\lambda}, x):= \liminf_{\ell \downarrow 0} \frac{\mu_{\lambda}(L_{\lambda}^{\ell}(x))}{\frac{\sqrt{3}}{4}\ell^2}\]
(an analogue of the first formula on \cite[p. 233]{PeSo96}).  The argument from \cite{PeSo96} is likely to yield 
\[\int_{2/3}^{1}\iint_{T_{\lambda}} \underline{D}(\mu_{\lambda}, x) \:d\mu_{\lambda}(x)\:d\lambda < \infty.\]
From that we could conclude that $\lambda \mapsto \underline{D}(\mu_{\lambda}, x)$ is finite for a.e. $\lambda \in (2/3, 1)$, putting the Radon-Nikodym derivative of $\mu_{\lambda}$ with respect to two-dimensional Lebesgue measure in $L^2$ for a.e. $\lambda$.
\end{enumerate}

\subsection{The nature of the overlaps and induced systems}\label{natureoverlap}

\bigskip

As we noted before, the nature of the overlaps changes at the value 
$\lambda = (\sqrt{5}-1)/2$. Here we refer to overlaps of monomials in the
$\tau_i$'s of degree $n = 1, 2, \ldots$ applied to the initial
triangle $T = T_{\lambda}$ as $\tau^n(T)$.   Let $\textbf{ov}(\tau^n(T))$ denote overlaps at level $n$---
for example, 
\[ \textbf{ov}(\tau^1(T)) = \Bigl( \tau_0(T)\cap\tau_1(T)\Bigr) \cup \Bigl( \tau_0(T)\cap\tau_2(T)\Bigr) \cup \Bigl( \tau_1(T)\cap\tau_2(T)\Bigr).\]

\begin{figure}
\setlength{\unitlength}{1bp}
\bigskip
\bigskip
\begin{picture}(324,162)(0,-45)
\put(86,0){\includegraphics[width=153\unitlength]{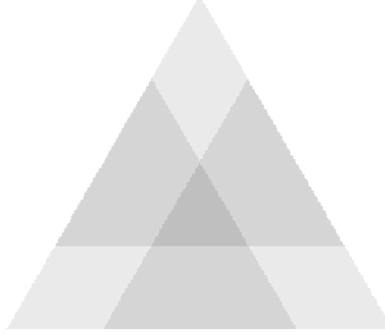}}
\end{picture}
\caption{The first iteration of $X_{\lambda}$ for $\lambda = \frac{3}{4}$.  In this case, there are no gaps.}
\label{FigSie.2}
\end{figure}

\begin{enumerate}[(i)]
\item\label{regularSierpinski}\textbf{The Sierpinski Gasket.  }
When $\lambda = 1/2$, the resulting fractal $X_{\lambda}$ is the Sierpinski gasket, and the essential overlap is a set of Lebesgue measure zero \cite{Str06}.
\item\label{essentialbutsimpleoverlap}\textbf{Simple Overlap.  }
When $\lambda\in(1/2, (\sqrt{5}-1)/2)$, we have
\[ \textbf{ov}(\tau^n(T)) \cap \textbf{ov}(\tau^{n+1}(T)) = \varnothing.\]
We call this type of overlap ``overlap of multiplicity one'' or ``simple overlap.''  When simple overlap occurs, the subset of $X_{\lambda}$ which consists of the overlaps is itself an IFS.  So, an IFS with essential but simple overlap induces a new IFS with non-essential overlap.  See Figure \ref{FigSie.4}.

\item \label{essentialsimple}\textbf{The Golden 2D Fractal.  }
Let $\lambda = (\sqrt{5}-1)/2$, as in Figure \ref{FigSie.1}.  We see $\textbf{ov}(\tau^1(T))$ as the  dark shaded triangles in Figure \ref{FigSie.1}, picture (a).  As we move from (a) to (b) in Figure \ref{FigSie.1}, we see that successive overlap sets intersect at vertices: 
$\textbf{ov}(\tau^1(T)) \cap \textbf{ov}(\tau^2(T))$ is non-empty.  In fact, $\textbf{ov}(\tau^1(T))$ consists of three triangles, and  $\textbf{ov}(\tau^2(T))$ of $9$.  Each triangle in $\textbf{ov}(\tau^1(T))$ shares a vertex with one or two from $\textbf{ov}(\tau^2(T))$, and the double-sharing happens for the interior triangles from $\textbf{ov}(\tau^2(T))$.  This pattern continues, so that at each step of the iteration, each vertex of a triangle in $\textbf{ov}(\tau^n(T))$ coincides with some vertex in a triangle from $\textbf{ov}(\tau^{n+1}(T))$.   
We therefore have an induced system which forms a graph with edges and vertices, with each vertex from $\textbf{ov}(\tau^n(T))$ connecting to one or two vertices from $\textbf{ov}(\tau^{n+1}(T))$.   

In the following when we refer to disjoint pairs of triangles, we will
mean ``disjointness of the respective interiors," thus allowing the sharing
of vertices.  The triangles in $\textbf{ov}(\tau^{n}(T))$ are disjoint from all the triangles in $\textbf{ov}(\tau^{n+1}(T))$, but triangles from $\textbf{ov}(\tau^{n+2}(T))$ may be contained in triangles from $\textbf{ov}(\tau^{n}(T))$.  In fact, a triangle from $\textbf{ov}(\tau^{n+2}(T))$ is either disjoint from triangles in the set $\textbf{ov}(\tau^{n}(T))$, or that triangle is contained in a unique triangle from $\textbf{ov}(\tau^{n}(T))$.

We can formalize the overlap between iterations by defining a set operation $\textrm{OV}$ which takes a set $S$ to the set 
\[ \textrm{OV}(S):=\Bigl( \tau_0(S)\cap\tau_1(S)\Bigr) \cup \Bigl( \tau_0(S)\cap\tau_2(S)\Bigr) \cup \Bigl( \tau_1(S)\cap\tau_2(S)\Bigr).\]%
We have already seen that 
\[ \textrm{OV}(T_{\lambda}) = \textbf{ov}(\tau^1(T)),\]
and we also have 
\[ \textrm{OV}(\textbf{ov}(\tau^1(T))) = \textbf{ov}(\tau^2(T)).\]
Suppose $\xi = (i_1, i_2, \ldots i_n)\in \{0, 1, 2\}^n$---that is, $\xi$ is a multi-index of length $n$ each of whose components is $0$, $1$, or $2$.  We can use $\xi$ to keep track of the monomials in the $\tau_i$'s which we mentioned above:
\[ \tau_{\xi}(x) : = \tau_{i_1}\tau_{i_2} \cdots \tau_{i_n}(x) = \lambda^n x + \lambda^n u_{i_1} + \lambda ^{n-1}u_{i_2} + \ldots \lambda u_{i_n}.\]  Then
\[ \textbf{ov}(\tau^{n+1}(T)) = \Bigl\{ \textrm{OV}(\tau_{\xi}(T)) : \xi \in \{0, 1, 2\}^n\Bigr\}.\]
\item \label{essentialmultiplicity}\textbf{The Residual Interval.  }
When $(\sqrt{5}-1)/2 < \lambda < 2/3$, the set $\textbf{ov}(\tau^{n}(T)) \cap \textbf{ov}(\tau^{n+1}(T))$ no longer consists of discrete points---for each $n$,  $\textbf{ov}(\tau^{n}(T)) \cap \textbf{ov}(\tau^{n+1}(T))$  is uncountable.  See, for example, Figure \ref{FigSie.3}, where $\lambda = 13/20$.  To be more precise, if we  rescale the initial triangle $T$ so that the length of each of its sides is $b = \lambda/(1-\lambda)$, then for each $n$, the set $\textbf{ov}(\tau^n(T))$ consists of $3^n$ disjoint triangles, each of side length $\lambda^n$.
\item\label{closinggap} \textbf{The Closing of the Gap.  }When $\lambda = 2/3$, there is still overlap with multiplicity, but there are no longer gaps at each iteration.  This pattern continues for $\lambda > 2/3$.  See Figure \ref{FigSie.3}, which shows a $\lambda$ value slighly less than $2/3$ and Figure \ref{FigSie.2}, which illustrates no gaps ($\lambda = 3/4$).
\end{enumerate}

\bigskip

\noindent\textsc{Summary of conclusions for 2D:  }

\bigskip

We have sketched features that come out differently for ``IFS
 overlap-Sierpinski fractals,'' stressing differences that arise when
passing from 1D examples where $\lambda \in (\frac{1}{2}, 1)$  to our analogous 2D attractors.   Specifically, in the 2D case, for the range of values of $\lambda$, there are five separate cases for scaling numbers $\lambda$ of interest,
 illustrating overlap features:

\begin{enumerate}[(i)]
\item  $\lambda\in (1/2, (\sqrt{5}-1)/2)$:   
$\textbf{ov}(\tau^{n}(T))\cap \textbf{ov}\tau^{n+1}(T)) = \varnothing$ (see Figure \ref{FigSie.4})
\item $\lambda = (\sqrt{5}-1)/2$:   $\textbf{ov}(\tau^{n}(T))\cap \textbf{ov}(\tau^{n+1}(T))$ consists of vertices (see Figure \ref{FigSie.1}); simple overlap; central gaps
\item $\lambda \in (\sqrt{5}-1)/2, 2/3)$:  $\textbf{ov}(\tau^{n}(T))\cap \textbf{ov}(\tau^{n+1}(T))$ consists of $3^n$ disjoint triangles; overlap with multiplicity; central gaps (see Figure \ref{FigSie.3} for a Sierpinski figure whose gaps are very small)
\item $\lambda = 2/3$:  central gaps close at $\lambda = 2/3$; overlap with multiplicity 
\item $\lambda\in (2/3 , 1)$:  no central gaps; overlap with multiplicity (see Figure \ref{FigSie.2}).
\end{enumerate}


\section{\label{Gen}Conclusions (the general case)}

We now return to the general case of IFSs with essential overlap. In this
case, the size of the overlap can nicely be expressed in terms of the column
isometry from Definition \ref{DefMul.ColIso}. To recall the setting, we begin
with the proof of Theorem \ref{ThmCon.EssOve} that was postponed.

\begin{proof}
[Proof of Theorem \textup{\ref{ThmCon.EssOve}}]A system of measurable
endomorphisms $\tau_{1},\dots,\tau_{N}$ in a finite measure space $\left(
X,\mathcal{B},\mu\right)  $ is given, and it is assumed that $\mu\left(
X\right)  =1$, and that%
\begin{equation}
\mu=\frac{1}{N}\sum_{i=1}^{N}\mu\circ\tau_{i}^{-1}, \label{eqOve.7}%
\end{equation}
i.e., that $\mu$ is a $\left(  \tau_{i}\right)  $-equilibrium measure. It then
follows from Proposition \ref{ProMul.EquMea} that the operators $F_{i}\colon
f\mapsto\frac{1}{\sqrt{N}}f\circ\tau_{i}$ define a column isometry, i.e., that
$\mathbb{F}=\left(
\begin{smallmatrix}
F_{1}\\
\vdots\\
F_{N}%
\end{smallmatrix}
\right)  $ satisfies $\mathbb{F}^{\ast}\mathbb{F}=I_{L^{2}\left(  \mu\right)
}$.

Note that this identity spells out to%
\begin{equation}
\sum_{i=1}^{N}F_{i}^{\ast}F_{i}=I_{L^{2}\left(  \mu\right)  },
\label{eqOve.poundA1}%
\end{equation}
but that in general, the \emph{individual} operators $F_{i}^{\ast}F_{i}$ are
not projections. We have the lemma:\renewcommand{\qed}{\relax}
\end{proof}

\begin{lemma}
\label{LemOve.MulOpe}Let $\left(  \tau_{i}\right)  _{i=1}^{N}$, $\mu$,
$\mathbb{F}=\left(  F_{i}\right)  _{i=1}^{N}$ be as above, i.e., $\mathbb{F}$
is a column isometry $L^{2}\left(  \mu\right)  \rightarrow L^{2}\left(
\mu\right)  _{N}$. Let $\varphi_{i}:=\frac{d\mu\circ\tau_{i}^{-1}}{d\mu}$.
Then%
\begin{equation}
F_{i}^{\ast}F_{i}=\frac{1}{N}M_{\varphi_{i}} \label{eqOve.poundA2}%
\end{equation}
where $M_{\varphi_{i}}$ is the multiplication operator $f\mapsto\varphi_{i}f$
in $L^{2}\left(  \mu\right)  $.
\end{lemma}

\begin{proof}
The result follows essentially from the argument in Lemma \ref{LemMul.AdjForm}
above. By that argument we may pass to partitions of $X$. Let $i$ be given,
fixed; and, following Lemma \ref{LemMul.AdjForm}, pass to a subset $E$ in $X$
such that there is a measurable mapping $\sigma_{E}\colon\tau_{i}\left(
E\right)  \rightarrow E$ with%
\begin{equation}
\sigma_{E}\circ\tau_{i}|_{E}=\operatorname{id}_{E}; \label{eqOve.poundA3}%
\end{equation}
see (\ref{eqMul.12}).

It follows that%
\[
F_{i}^{\ast}F_{i}f|_{E}=\frac{1}{N}\varphi_{i}f|_{E}%
\]
for all $f\in L^{2}\left(  \mu\right)  $. But the set $E$ is part of a finite
measurable partition of $X$, so the desired conclusion (\ref{eqOve.poundA2})
holds on $X$.
\end{proof}

\begin{proof}
[Proof of Theorem \textup{\ref{ThmCon.EssOve}} continued]Our assertion is that
$\mathbb{FF}^{\ast}=\left(  F_{i}F_{j}^{\ast}\right)  _{i,j=1}^{N}$ is the
identity operator in $L^{2}\left(  \mu\right)  _{N}$ if and only if the system
is essential non-overlap.

Setting $\varphi_{i}:=\frac{d\mu\circ\tau_{i}^{-1}}{d\mu}$ and using Lemma
\ref{LemMul.AdjForm} we show that there are measurable and invertible point
transformations $T_{i,j}\colon X\rightarrow X$, $i,j=1,\dots,N$, such that
$T_{i,i}=\operatorname{id}_{X}$, $1\leq i\leq N$, and%
\begin{equation}
F_{i}F_{j}^{\ast}=\frac{1}{N}\left(  \varphi_{j}\circ\tau_{i}\right)  T_{i,j}.
\label{eqOve.8}%
\end{equation}
But, for each $i$, the function $\varphi_{i}$ is supported on $\tau_{i}\left(
X\right)  $. So if%
\begin{equation}
F_{i}F_{j}^{\ast}=\delta_{i,j}I_{L^{2}\left(  \mu\right)  }, \label{eqOve.9}%
\end{equation}
then%
\begin{align*}
\varphi_{i}  &  \equiv N &  &  \mu\text{-}\mathrm{a.e.}\text{ on }\tau
_{i}\left(  X\right) \\%
\intertext{and}%
\varphi_{i}  &  \equiv0 &  &  \mu\text{-}\mathrm{a.e.}\text{ on }%
X\setminus\tau_{i}\left(  X\right)  =\bigcup_{k\neq i}\tau_{k}\left(
X\right)  .
\end{align*}
The conclusion of the theorem is immediate from this; and we get the following corollary.
\end{proof}

\begin{corollary}
\label{CorOve.Onto}Let the IFS $\left(  \tau_{i}\right)  _{i=1}^{N}$ be as
specified in Theorem \textup{\ref{ThmCon.EssOve}} above, and let
$\mathbb{F}\colon L^{2}\left(  \mu\right)  \rightarrow L^{2}\left(
\mu\right)  _{N}$ be the corresponding column isometry, with $F_{i}\colon
f\mapsto\frac{1}{\sqrt{N}}f\circ\tau_{i}$. Then $\mathbb{F}$ maps onto
$L^{2}\left(  \mu\right)  _{N}$ if and only
\begin{equation}
\left(  F_{i}^{\ast}f\right)  =\sqrt{N}\chi_{\tau_{i}\left(  X\right)
}\left(  x\right)  f\left(  \sigma_{i}\left(  x\right)  \right)  \qquad
\mu\text{-}\mathrm{a.e.}\;x\in X. \label{eqOve.10}%
\end{equation}

\end{corollary}

(Here the endomorphisms $\sigma_{i}\colon X\rightarrow X$ are specified in
Lemma \ref{LemMul.AdjForm}, and in particular $\sigma_{i}\circ\tau
_{i}=\operatorname{id}_{X}$, $1\leq i\leq N$.)

\begin{theorem}
\label{ThmGen.IsoExt}Let $N\in\mathbb{N}$, $N\geq2$, be given, and let
$\left(  \tau_{i}\right)  _{i\in\mathbb{Z}_{N}}$ be a contractive IFS in a
complete metric space. let $\left(  X,\mu\right)  $ be the Hutchinson data;
see Definition \textup{\ref{DefCon.LimMea}}. Let $P$ \textup{(}$=P_{1/N}%
$\textup{)} be the Bernoulli measure on $\Omega=\prod_{1}^{\infty}%
\mathbb{Z}_{N}=\mathbb{Z}_{N}^{\mathbb{N}}$; see Corollary
\textup{\ref{CorCon.EquiMe}}. Let $\pi\colon\Omega\rightarrow X$ be the
encoding mapping of Lemma \textup{\ref{LemCon.CodCon}}. Set%
\begin{align}
F_{i}f &  :=\frac{1}{\sqrt{N}}f\circ\tau_{i}\text{\qquad for }f\in
L^{2}\left(  X,\mu\right)  \label{eqGen.9}\\[1pt]%
\intertext{and}%
S_{i}^{\ast}\psi &  :=\frac{1}{\sqrt{N}}\psi\circ\sigma_{i}\text{\qquad for
}\psi\in L^{2}\left(  \Omega,P\right)  ,\label{eqGen.10}%
\end{align}
where $\sigma_{i}$ denotes the shift map of \textup{(\ref{eqCon.7})}.

\setcounter{enumi}{0}\renewcommand{\theenumi}{\alph{enumi}}\refstepcounter
{enumi}\label{ThmGen.IsoExt(1)}\textup{(\ref{ThmGen.IsoExt(1)})} Then the
operator $V\colon L^{2}\left(  X,\mu\right)  \rightarrow L^{2}\left(
\Omega,P\right)  $ given by%
\begin{equation}
Vf=f\circ\pi\label{eqGen.11}%
\end{equation}
is isometric.

\refstepcounter{enumi}\label{ThmGen.IsoExt(2)}\textup{(\ref{ThmGen.IsoExt(2)}%
)} The following intertwining relations hold:%
\begin{equation}
VF_{i}=S_{i}^{\ast}V,\qquad i\in\mathbb{Z}_{N}. \label{eqGen.12}%
\end{equation}

\refstepcounter{enumi}\label{ThmGen.IsoExt(3)}\textup{(\ref{ThmGen.IsoExt(3)}%
)} The isometric extension $L^{2}\left(  X,\mu\right)  \hookrightarrow
L^{2}\left(  \Omega,P\right)  $ of the $\left(  F_{i}\right)  $-relations is
minimal in the sense that $L^{2}\left(  \Omega,P\right)  $ is the closure of%
\begin{equation}
\bigcup_{n}\bigcup_{i_{1}i_{2}\dots i_{n}}S_{i_{1}}S_{i_{2}}\cdots S_{i_{n}%
}VL^{2}\left(  X,\mu\right)  . \label{eqGen.13}%
\end{equation}

\end{theorem}

\begin{proof}
\emph{Ad} (\ref{ThmGen.IsoExt(1)})--(\ref{ThmGen.IsoExt(2)}): Let $f\in
L^{2}\left(  X,\mu\right)  $, and let $\left\Vert \,\cdot\,\right\Vert _{\mu}$
and $\left\Vert \,\cdot\,\right\Vert _{P}$ denote the respective $L^{2}$-norms
in $L^{2}\left(  \mu\right)  $ and $L^{2}\left(  P\right)  $.
Then\settowidth{\equallabelwidth}{$\scriptstyle\text{by (\ref{eqCon.13})}$}%
\begin{align*}
\left\Vert Vf\right\Vert _{P}^{2} &
\underset{\makebox[\equallabelwidth]{{}}}{=}\int_{\Omega}\left\vert f\circ
\pi\right\vert ^{2}\,dP\\
&  \underset{\makebox[\equallabelwidth]{{}}}{=}\int_{X}\left\vert f\right\vert
^{2}\,d\left(  P\circ\pi^{-1}\right)  \\
&  \underset{\text{by (\ref{eqCon.13})}}{=}\int_{X}\left\vert f\right\vert
^{2}\,d\mu=\left\Vert f\right\Vert _{\mu}^{2}.
\end{align*}
Moreover,\settowidth{\equallabelwidth}{$\scriptstyle\text{by (\ref{eqCon.11})}$}%
\begin{align*}
VF_{i}f &  \underset{\makebox[\equallabelwidth]{{}}}{=}\left(  F_{i}f\right)
\circ\pi\\
&  \underset{\makebox[\equallabelwidth]{{}}}{=}\frac{1}{\sqrt{N}}f\circ
\tau_{i}\circ\pi\\
&  \underset{\text{by (\ref{eqCon.11})}}{=}\frac{1}{\sqrt{N}}f\circ\pi
\circ\sigma_{i}\\
&  \underset{\makebox[\equallabelwidth]{{}}}{=}S_{i}^{\ast}Vf,
\end{align*}
which is assertion (\ref{ThmGen.IsoExt(2)}).

\emph{Ad} (\ref{ThmGen.IsoExt(3)}): Let $\psi\in L^{2}\left(  \Omega,P\right)
$, and let $\left\langle \,\cdot\mid\cdot\,\right\rangle _{\mu}$ and
$\left\langle \,\cdot\mid\cdot\,\right\rangle _{P}$ denote the respective
Hilbert inner products of $L^{2}\left(  \mu\right)  $ and $L^{2}\left(
P\right)  $. To show that the space in (\ref{eqGen.13}) is dense in
$L^{2}\left(  P\right)  $, suppose%
\begin{equation}
0=\left\langle \,S_{i_{1}}\cdots S_{i_{n}}Vf\mid\psi\,\right\rangle
_{P}\label{eqGen.14}%
\end{equation}
for all $n$, all multi-indices $\left(  i_{1}\dots i_{n}\right)  $, and all
$f\in L^{2}\left(  \mu\right)  $. We will prove that then $\psi=0$.

When $\left(  i_{1}\dots i_{n}\right)  $ is fixed, we denote the cylinder set
in $\Omega$ by%
\begin{equation}
C\left(  i_{1},\dots,i_{n}\right)  =\left\{  \,\omega\in\Omega\mid\omega
_{j}=i_{j},\;1\leq j\leq n\,\right\}  .\label{eqGen.15}%
\end{equation}
Using now (\ref{eqOve.10}) in Corollary \ref{CorOve.Onto} on $\Omega$, we get
\[
S_{i_{n}}^{\ast}\cdots S_{i_{1}}^{\ast}\psi=N^{-n/2}\psi\circ\sigma_{i_{1}%
}\circ\dots\circ\sigma_{i_{n}}.
\]
Substitution into (\ref{eqGen.14}) yields
\[
\int_{\Omega}\chi_{C\left(  i_{1},\dots,i_{n}\right)  }\psi\,dP=0.
\]
We used the fact that (\ref{eqGen.14}) holds for all $f\in L^{2}\left(
\mu\right)  $. But the indicator functions $\chi_{C\left(  i_{1},\dots
,i_{n}\right)  }$ span a dense subspace in $L^{2}\left(  \Omega,P\right)  $
when $n$ varies, and all finite words of length $n$ are used. We conclude that
$\psi=0$, and therefore that the space in (\ref{eqGen.13}) is dense in
$L^{2}\left(  \Omega,P\right)  $.
\end{proof}

\begin{remark}
\label{RemGen.4}Note that by \textup{(\ref{eqGen.12})} the space in
\textup{(\ref{eqGen.13})}, part \textup{(\ref{ThmGen.IsoExt(3)})} of the
theorem, is invariant under the operators $S_{i}^{\ast}$.
\end{remark}

\begin{acknowledgements}
We are pleased to thank Dorin Dutkay for helpful conversations. We thank Brian
Treadway for excellent typesetting, and for producing the graphics.
\end{acknowledgements}

\bibliographystyle{alpha}

\end{document}